\documentclass[a4paper,11pt]{article}
\usepackage{jcappub} % for details on the use of the package, please
\usepackage{cite}
\usepackage{graphicx}
\usepackage{enumerate}
\usepackage{subfig}
\usepackage{amsmath}
\usepackage[T1]{fontenc} % if needed
\usepackage[utf8]{inputenc}
\usepackage{hyperref}
\usepackage[T1]{fontenc}

\usepackage{slashed} 
\usepackage[mathscr]{eucal}
\usepackage[Symbolsmallscale]{upgreek}
\usepackage{isotope} %for chemical symbol of Tritium
\title{\boldmath 
Shedding New Light on Sterile Neutrinos from XENON1T Experiment
}

\author[a,b]{Soroush Shakeri,
}
\author[c]{Fazlollah Hajkarim,}
\author[d]{She-Sheng Xue}

\affiliation[a]{Department of Physics, Isfahan University of Technology, Isfahan 84156-83111, Iran}
\affiliation[b]{ICRANet-Isfahan, Isfahan University of Technology, 84156-83111, Iran}
\affiliation[c]{Institut f{\"u}r  Theoretische  Physik,  Goethe  Universit{\"a}t, Max  von  Laue  Stra{\ss}e  1, \\  D-60438  Frankfurt  am  Main,  Germany}
\affiliation[d]{ICRANet Piazzale della Repubblica, 10 -65122, Pescara, Italy, \\ Physics Department, University of Rome La Sapienza, \\ P.le Aldo Moro 5, I–00185 Rome, Italy
}

\emailAdd{s.shakeri@iut.ac.ir}
\emailAdd{hajkarim@th.physik.uni-frankfurt.de}
\emailAdd{xue@icra.it}

\abstract{The XENON1T  collaboration recently reported the excess of events from recoil electrons, possibly giving an insight into new area beyond the Standard Model (SM)  of particle physics. We try to explain this excess by considering  effective interactions between the sterile neutrinos and the SM particles. 
In this paper, we present an effective model based on  one-particle-irreducible interaction vertices at low energies that are induced from the SM 
gauge symmetric four-fermion operators at high energies. The effective interaction strength  is constrained by the SM precision measurements, astrophysical and cosmological observations. We introduce a novel effective electromagnetic interaction between sterile neutrinos and SM neutrinos, which can successfully explain the XENON1T event rate through inelastic scattering of the sterile neutrino dark matter from Xenon electrons.  
We find that sterile neutrinos with masses around  $90$ keV and specific effective coupling can fit well with the XENON1T data where the best fit points preserving DM constraints and possibly describe the anomalies in other experiments. 
}

\begin{document}
\maketitle
\flushbottom

%%%%%%%%%%%%%%%%%%%%%%%%%%%%%%%%%%%%%%%%%%%%%%%%%%%%%%%%%%%%%%%%%%%%%%%%%%%%%%%%%%%%%%%%%%%%%%%%%%%%%%%%%%%%%%%%%%%%%%%%%%%%
%%%%%%%%%%%%%%%%%%%%%%%%%%%%%%%%%%%%%%%%%%%%%%%%%%%%%%%%%%%%%%%%%%%%%%%%%%%%%%%%%%%%%%%%%%%%%%%%%%%%%%%%%%%%%%%%%%%%%%%%%%%%
%%%%%%%%%%%%%%%%%%%%%%%%%%%%%%%%%%%%%%%%%%%%%%%%%%%%%%%%%%%%%%%%%%%%%%%%%%%%%%%%%%%%%%%%%%%%%%%%%%%%%%%%%%%%%%%%%%%%%%%%%%%%
%
\section{Introduction}
\label{sec:intro}

Recently the XENON1T experiment has reported an excess of $\mathcal{O}(\text{keV})$ electronic recoil events over known Standard Model (SM) background with a statistical significance of $\sim 3 \, \sigma$ \cite{Aprile:2020tmw}. The excess is peaked around the 2 and 3 keV energy bins in a 1-7 keV recoil energy window. The XENON collaboration has been unable to exclude the $\beta$ decay of Tritium as a background responsible for the surplus events. Even a trace amount of fewer than three $\isotope[3]{H}$ atoms per kg of XENON is enough to fit the excess with a $3.2 \, \sigma$ significance \cite{XENON1T_talk}. Despite viable SM explanations regarding unresolved backgrounds \cite{Szydagis:2020isq,Bhattacherjee:2020qmv, Robinson:2020gfu}, it remains a reasonable possibility that this excess is a hint of new physics beyond the SM. 
The experimental analysis takes the latter into account and considers the possibility that the excess can be due to a solar axion or a solar neutrino with a sizeable magnetic moment being absorbed by target electrons. However, both of these scenarios are in tension with stellar cooling constraints \cite{DiLuzio:2020jjp,Gao:2020wer,Dent:2020jhf,2019arXiv191010568A,Viaux:2013lha,Athron:2020maw}. 

In  the SM, the neutrino-electron scattering as a source of  electron recoils, generically originates from the exchange of the SM gauge bosons $W^{\pm}$ and $Z_0$ \cite{Mohapatra:1998rq,Itzykson:1980rh,Tomalak:2019ibg}. Nonzero magnetic moments of neutrinos can also provide some additional contributions 
 to the low energy neutrino-electron  scattering via electromagnetic (EM) interactions \cite{Broggini:2012df,PhysRev.41.763,deGouvea:2006hfo,Fujikawa:1980yx,Bhattacharya:2002aj}. However, the precise value of the neutrino magnetic moment is unknown and it is predicted to be very small in the SM, proportional to the neutrino masses for Dirac neutrinos \cite{Vogel:1989iv,Fujikawa:1980yx}. 
 The nonstandard interactions of the SM neutrinos induce large neutrino magnetic moments which can account for the XENON1T excess events \cite{Khan:2020vaf,Babu:2020ivd,Chala:2020pbn,Amaral:2020tga,Miranda:2020kwy,Arcadi:2020zni}. In order to avoid
large corrections to neutrino masses induced by  large neutrino magnetic moments, some new symmetries can be introduced \cite{Babu:2020ivd,Lindner:2017uvt}.
Moreover, there are many theoretical attempts to interpret the  XENON1T signal assuming new physics  beyond the SM such as, boosted dark matter \cite{Fornal:2020npv,Jho:2020sku,Alhazmi:2020fju,Davoudiasl:2020ypv,Dey:2020sai,McKeen:2020vpf}, dark radiation \cite{Alonso-Alvarez:2020cdv,Chiang:2020hgb }, anomalous magnetic moment of muon \cite{Amaral:2020tga,Borah:2020jzi},  inelastic DM-electron scattering \cite{Harigaya:2020ckz,Bell:2020bes,Lee:2020wmh,Bramante:2020zos,Smirnov:2020zwf,Choudhury:2020xui}, axion-like  DM \cite{Takahashi:2020bpq,Athron:2020maw}.
Besides, there are some proposals using a gauged $U(1)_{X}$ extention of the SM, where a gauge boson $Z'$ contributes to dark matter (DM) \cite{Bally:2020yid,Okada:2020evk,Lindner:2020kko,Boehm:2020ltd,AristizabalSierra:2020edu,Choi:2020udy,Choi:2020kch,Kim:2020aua,Baek:2020owl}. A detailed study of electron-neutrino scattering in  the framework of $U(1)_{X}$ was presented in \cite{Lindner:2018kjo}.

In this article, we argue that the sterile neutrino DM  can account for the low-energy excess at XENON1T, while evading cosmological and astrophysical bounds.  Right-handed sterile neutrinos are introduced to resolve different theoretical problems in the SM and simultaneously can be served  
as a viable DM  candidates, see for example Refs.~\cite{Xue:1997tz,Xue:2001he,Drewes:2013gca}.
Sterile neutrinos as a warm DM with masses at  keV scale  may constitute all or a part of galactic DM halo \cite{Dodelson:1993je,Boyarsky:2018tvu,Boyarsky:2012rt,PhysRevLett.72.17,Kusenko:2009up,Adhikari:2016bei}. They can satisfy the bounds from structure formation and the free streaming length of DM at early epochs \cite{Adhikari:2016bei,Drewes:2013gca,Boyarsky:2018tvu}, and also explain the deviation of number of effective neutrinos measured from cosmic microwave background (CMB)  \cite{Gelmini:2019esj,Drewes:2013gca}.
Moreover, the presence of sterile neutrinos might explain the baryon asymmetry of the Universe  \cite{Asaka:2005pn,Shaposhnikov:2008pf}. 
As soon as finding any evidences for the sterile neutrino DM and its properties one  can gain insightful information about the thermal history of the early Universe and the production mechanism of these particles  \cite{Gelmini:2004ah,Drewes:2013gca}.

We present an effective  model based on the fundamental symmetries and particle content of the SM, in order to describe the relevant effective interactions of right-handed sterile neutrinos with SM particles at low energies. In this scenario three massive sterile neutrinos are introduced, the SM gauge symmetric four-fermion interactions giving rise to new effective interactions between sterile neutrinos and SM gauge bosons.
Using these effective interactions, we calculate the scattering cross sections between sterile neutrinos and electrons at the XENON detector. We will show that inelastic scattering of  sterile neutrino DM from Xenon electrons is  able to successfully reconstruct the XENON1T event rate.
The interaction of sterile neutrinos with electrons for explaining the XENON1T excess has been recently studied in  ~\cite{Shoemaker:2018vii,Ge:2020jfn}, where  an incoming flux of solar neutrinos  up scatters to  an outgoing state of sterile neutrinos. In  contrast to the case of incoming sterile neutrinos, their results  do not rely on the nature of sterile neutrinos as DM particles.

This paper is organized as follows. In the next section  we explain the framework for sterile neutrino interactions with other SM particles. Then we calculate the possible interaction with electrons (Sec.~\ref{sec:scat}) which might lead to the excess of events for XENON1T electron recoil at low energy. In Sec.~\ref{sec:recoil} we compute the recoil energy of electrons due to sterile neutrino interactions either being in the initial state or in the final state. Moreover, we discuss the possible bounds on the parameters of our effective model. Finally, we discuss our results and conclude in Sec.~\ref{sec:conc}.

%%%%%%%%%%%%%%%%%%%%%%%%%%%%%%%%%%%%%%%%%%%%%%%%%%%%%%%%%%%%%%%%%%%%%%%%%%%%%%%%%%%%%%%%%%%%%%%%%%%%%%%%%%%%%%%%%%%%%%%%%%%%
%%%%%%%%%%%%%%%%%%%%%%%%%%%%%%%%%%%%%%%%%%%%%%%%%%%%%%%%%%%%%%%%%%%%%%%%%%%%%%%%%%%%%%%%%%%%%%%%%%%%%%%%%%%%%%%%%%%%%%%%%%%%
%%%%%%%%%%%%%%%%%%%%%%%%%%%%%%%%%%%%%%%%%%%%%%%%%%%%%%%%%%%%%%%%%%%%%%%%%%%%%%%%%%%%%%%%%%%%%%%%%%%%%%%%%%%%%%%%%%%%%%%%%%%%

\section{Sterile Neutrino and Charged Lepton Couplings}\label{ch2}

Chiral gauge symmetries  and  spontaneous/explicit breaking of these symmetries are key concepts to understand the hierarchy pattern of fermion masses in the SM, and furthermore play crucial role in developing  possible scenarios beyond the SM for the fundamental particle physics. As demonstrated by neutrino oscillation phenomenon \cite{Fukuda:1998mi},  apart from being Dirac or Majorana type,  neutrinos are massive particles. This fact can imply  the existence of the right-handed neutrinos and raises the questions of how chiral gauge symmetries have been broken to generate neutrino masses.

There are many possibilities  with various degree of complexity to capture such a new physics where more unknown particles with different masses, interaction,  spin and strength  might be included in the model. The seesaw mechanism is one of the most economical solution to neutrino mass problem which is implemented in three tree-level ideas so-called as type-I \cite{Minkowski:1977sc,Glashow:1979nm,GellMann:1980vs,Schechter:1980gr,PhysRevLett.44.912,Haghighat:2019rht}, type-II \cite{Cheng:1980qt} and
type-III  \cite{Foot:1988aq}. In some of the models in order to avoid  very small tree-level Yukawa term, new discrete symmetries or continuous global symmetries are assumed \cite{Ma:2014qra,Ma:2015mjd}. Another way is to extend  the SM with new gauge symmetries such as $U(1)_{B-L}$ \cite{PhysRevD.20.776,PhysRevLett.44.1316,Marshak:1979fm,Wetterich:1981bx}
, or $U(1)_{R}$ \cite{Gu:2007ug}, such symmetry extensions in the gauge group may lead to introducing several new beyond SM gauge bosons. Moreover left-right symmetric models with direct right-handed neutrino interactions and SM gauge group $SU(2)_{L}\times SU(2)_{R}\times U(1)$ are among the attractive extensions of the SM \cite{Marciano:1977wx,PhysRevD.11.2558,PhysRevD.12.1502}.

On the other hand, the theoretical inconsistency \cite{Nielsen:1980rz,Nielsen:1981xu,1981PhLB..105..219N,Nielsen:1991si} between the SM bilinear Lagrangian of the chiral gauged fermions and the natural UV cutoff of unknown dynamics or quantum gravity requires that an effective field theory possesses quadrilinear four-fermion interactions (operators) of the Nambu-Jona-Lasinio (NJL) type \cite{Nambu:1961fr} at high-energy scales.
On the basis of {\it only} SM gauge symmetries, four-fermion operators of SM left- and right-handed fermions $(\psi_{_L},\psi_{_R})$ in the charge sector ``$Q$'' and flavor family ``{\it f}'' are introduced \cite{Xue:2016dpl,Xue:2016txt,Xue:2015wha,Xue:2001he,Xue:2014opa,Xue:1997tz,Xue:1996fm}
\begin{eqnarray}\label{ffermi}
\sum_{f=1,2,3}G\, \Big[\bar\psi^{f}_{_L}\psi^{f}_{_R}\bar\psi^{f}_{_R} \psi^{f}_{_L}\Big]_{Q_i=0,-1,2/3,-1/3}\,,
\label{q1}
\end{eqnarray}
and three SM gauge-singlet right-handed neutrinos $\psi^{f}_{_R}=\nu_R^f$ \footnote{It is not excluded that they are all identical or related.} 
are in the sector $Q=0$, playing the role of warm dark matter.
There are two fixed points (scaling domains) for the effective coupling $G$ in (\ref{ffermi}). One is the strong coupling ultraviolet (UV) fixed point at characteristic energy TeV, where SM gauge symmetries are preserved by composite particles \cite{Xue:2016txt}. A phenomenological study of such composite particles at the LHC is recently presented in \cite{Leonardi:2018jzn}. Another one is the weak-coupling infrared (IR) 
fixed point at the electroweak scale $\sim 246$ GeV, 
where the low-energy SM model is realized. The SM gauge symmetries are spontaneously broken, top quark, $W^\pm$ and $Z^0$ gauge bosons acquire their masses \cite{Bardeen:1989ds}. The hierarchy massive spectra of SM Dirac fermions are resulted from 
the explicit symmetry breaking due to the family mixing, 
and the seesaw mechanism leads to Majorana neutrino masses \cite{Xue:2016dpl}.

The relevant feature of this model for the present  article is that in the IR low energy domain one-particle-irreducible (1PI) interacting vertices between
left- and right-handed fermions are induced by the four-fermion operators (\ref{ffermi}). These lead to additional right-handed currents coupling to SM bosons $W^\pm$ and $Z^0$ in the neutrino sector \cite{Xue:1997tz,Xue:1999xa} and the quark sector \cite{Xue:1996fm,Xue:2015wha}. These 1PI left-right mixing  vertices vanish at low energies, giving rise to the chiral (parity-violated) symmetries of SM. However, they do not vanish at high energies, implying that 
the parity symmetry could be restored at high energy scale
\cite{Xue:2001he,Xue:2013fla}. Recently, the effective operators of right-handed currents have been considered for studying LHC physics \cite{Alioli:2017ces}, the circular polarization of cosmic photons due to right-handed neutrino DM  candidate is also considered in \cite{Haghighat:2019rht}.

\begin{figure}
  \centering
    \subfloat{
    \centering
    \includegraphics[width=1.4in]{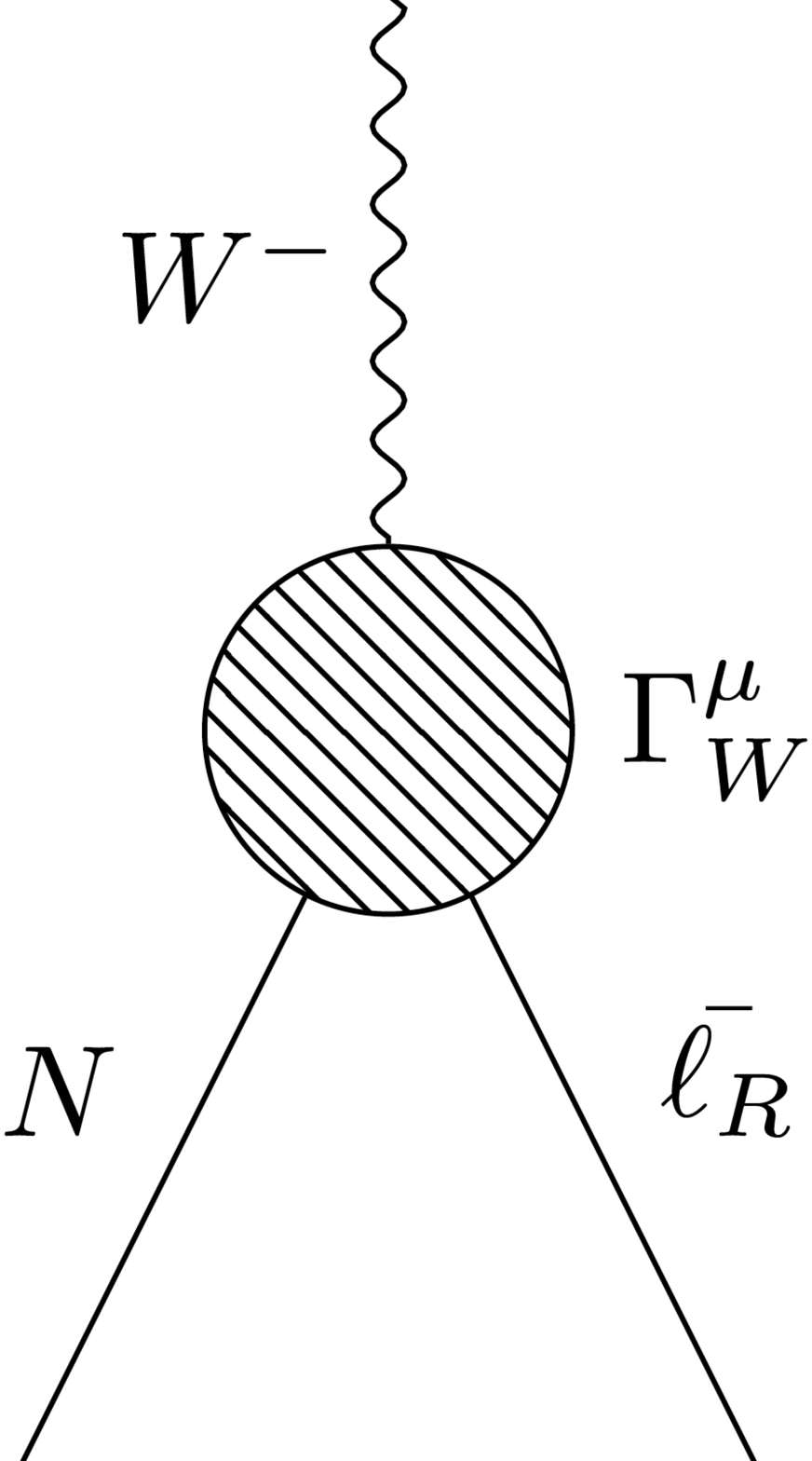} 
    \hspace{2.4cm}
    \includegraphics[width=1.4in]{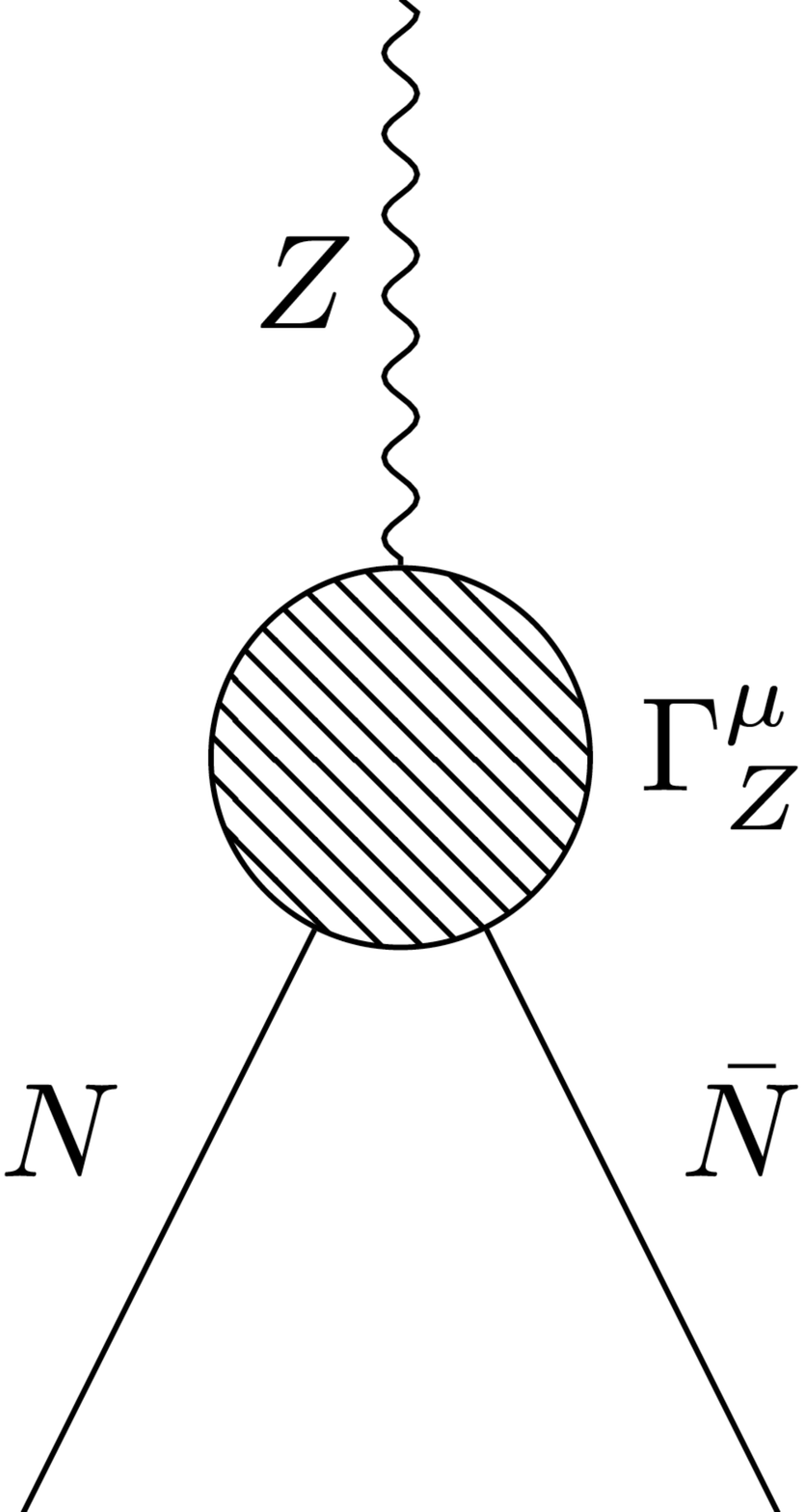}
    }
 \caption{The Feynman diagram of the effective right-handed current couplings in Eq.~(\ref{rhc}). Left: The effective vertex of sterile neutrino and lepton coupling to $W^{-}$ with the mixing matrix $[(U^\ell_R)^\dagger U^\nu_R]$. Right: The effective vertex of sterile neutrino and anti-sterile neutrino  coupling to $Z^0$. }
  \label{fig1}
\end{figure}

We consider right-handed sterile neutrinos as the DM candidate effectively couple to the SM gauge bosons via the right-handed charged and neutral current interactions as \cite{Xue:1997tz,Xue:1996fm,Xue:2001he}
\begin{eqnarray}
	\mathcal{L}&\supset & {\mathcal{G}_R}({g_w}/{\sqrt{2}})~\bar \ell_R\gamma^\mu\nu^\ell_RW^-_\mu +{\mathcal{F}_R}({g_w}/{\sqrt{2}})~\bar \nu^\ell_R\gamma^\mu\nu^\ell_R Z_\mu +	{\rm h.c.}\,
	\label{rhc0}
\end{eqnarray}
where $G_{F}/\sqrt{2}=g_{w}^{2}/8M_{W}^{2}$ and $\nu^\ell_R$ stands for the right-handed neutrino in the same family of the charged lepton $\bar\ell_R$, $\nu^\ell_R$ and $\ell_R$ can be considered as a right-handed doublet $(\nu^\ell_R, \ell_R)$ in  
gauge interacting basis. Summation over three lepton families $\ell$ is performed in Eq.~(\ref{rhc0}) and no flavour-changing-neutral-current (FCNC) interactions occur. The ${\mathcal{G}}_R$ and ${\mathcal{F}}_R$ represent effective coupling vertices, in the momentum space, given by (see Fig.~ \ref{fig1})
\begin{equation}\label{IP1}
\Gamma_{W}^{\mu}=i\frac{\mathcal{G}_{R}(p_1,p_2) }{\sqrt{2}}\gamma^{\mu}g_{w}P_{R}\,, \ \ \ \ \ \ \ \ \ \Gamma^{\mu}_{Z}=i\frac{\mathcal{F}_{R}(p_1,p_2)}{\sqrt{2}}\gamma^{\mu}g_{w}P_{R}\,.
\end{equation}
They are complex functions of momenta $(p_1,p_2)$, and should be suppressed at low energies, as required by SM precision measurements and cosmological constraints. 
At the electroweak scale
$v=(G_F\sqrt{2})^{-1/2}\approx 246$ GeV, $\mathcal{G}_{R}\approx\mathcal{F}_{R}
\propto (v/\Lambda_{\rm cut})^2$. The UV cutoff $\Lambda_{\rm cut}$ is the scale at which the four-fermion operators (\ref{ffermi}) representing unknown new physics become relevant.

In terms of mass eigenstates $( N^l_R,l_R)$, 
gauge eigenstates $\nu^\ell_R = (U^\nu_R)^{\ell l'}N^{l'}_R$ and $\ell_R= (U^\ell_R)^{\ell l'} l'_R$, where $U^{\nu}_R$ and $U^{\ell}_R$ are $3\times 3$ unitary matrices in family flavor space, the 1PI interactions (\ref{rhc0})  take the following form
\begin{eqnarray}
	\mathcal{L}&\supset &
{\mathcal{G}_R}({g_w}/{\sqrt{2}})~[(U^\ell_R)^\dagger U^\nu_R]^{ll'}\bar l_R\gamma^\mu N^{l'}_RW^-_\mu +{\mathcal{F}_R}({g_w}/{\sqrt{2}})~\bar N^l_R\gamma^\mu N^l_R Z_\mu +	{\rm h.c.}\,
	\label{rhc}
\end{eqnarray}
and flavor mixing matrix $[(U^\ell_R)^\dagger U^\nu_R]$ appears in charged current interaction, while neutral current one remains diagonal in lepton family flavor space. The off-diagonal elements of mixing matrix $[(U^\ell_R)^\dagger U^\nu_R]$ shows interactions between different flavour families through the charged current channel, for example $N_R^e-\tau_R$. 
 This in fact gives the flavour oscillations of the sterile neutrinos \cite{Xue:2016txt}.  The mixing matrix $[(U^\ell_R)^\dagger U^\nu_R]$ is not the Pontecorvo-Maki-Nakagawa-Sakata (PMNS) matrix $[(U^\ell_L)^\dagger U^\nu_L]$ in the sector of left-handed leptons and neutrinos. 
That is due to the transformations from the mass eigenstates $(l_L, \nu^l_L)$ to the gauge eigenstates $\nu^\ell_L = (U^\nu_L)^{\ell l'}\nu^{l'}_L$ and $\ell_L= (U^\ell_L)^{\ell l'} l'_L$.  
Since there is no information 
about the mixing matrix $[(U^\ell_R)^\dagger U^\nu_R]$, we  assume in this article that off-diagonal mixing elements are very small, namely $[(U^\ell_R)^\dagger U^\nu_R]^{ll'}\approx \delta^{ll'}$. It is a reasonable assumption due to the hierarchy structure of charged lepton masses and thus the diagonal element values of order of unity have been absorbed in the ${\mathcal G}_R$ coupling.

\begin{figure}
    \centering
    \includegraphics[width=2.5in]{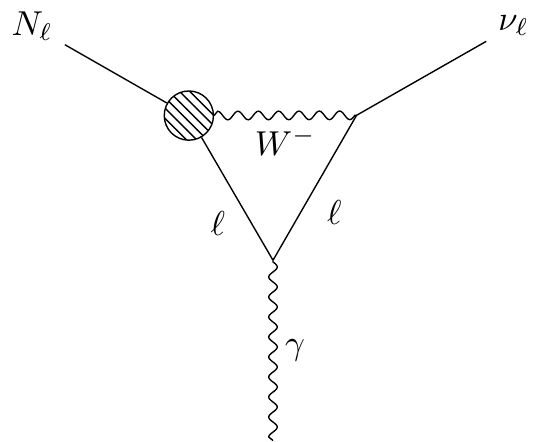}
    \caption{The effective electromagnetic (EM) interaction vertex of normal and sterile neutrinos. The PMNS mixing matrix  
    $(U^\nu_L U^\ell_L)$ associates to the SM coupling vertex $\bar\nu_L \gamma_\mu\ell_L W^\mu$.
    This effective vertex leads to the dominate   radiative decay of right-handed sterile neutrino $N_l\rightarrow \nu_l + \gamma$\,. }
    \label{enu2}
\end{figure}

 The sterile neutrinos via interaction (\ref{rhc}) at loop level can interact with photons which is given by Feynman diagram in Fig.~\ref{enu2}, where SM gauge boson $W$ and  charged lepton are present in the loop. In fact, this is an EM channel of SM neutrino and sterile neutrino interaction. It can be 
 represented as an effective operator  
\begin{equation}
 (U^\nu_L U^\ell_L)^{l l'} \bar\nu_L^l\Lambda^\mu_{l'} N_R^{l'} A_\mu + {\rm h.c.},
\label{effem0}
\end{equation} 
where $A^\mu$ is the electromagnetic field and $(U^\nu_L U^\ell_L)$ is the PMNS mixing matrix associating to the SM vertex $\bar \nu^l\gamma^\mu P_L\ell^l W_\mu$ in the loop. In the momentum space, the 1PI vertex $\Lambda_\mu$ is given by 
\begin{equation}
\Lambda_{l'}^\mu(q) =i\frac{e g_w^2 {\mathcal G}_Rm_{l'}}{16\pi^2}\Big[(C_0+2C_1)p_1^\mu+(C_0+2C_2)k_1^\mu\Big]\,.
\label{effem1}
\end{equation}
Here $p_{1}^{\mu}$ and $k_{1}^{\mu} $ denote the four-momenta of incoming sterile neutrino and outgoing SM neutrinos, respectively. The coefficients $C_0$, $C_1$ and $C_2$ are the three-point Passarino-Veltman functions \cite{Passarino:1978jh}, \begin{equation}
\label{loopfac3}
    C_i \equiv C_i (m_N^2, q^2, m_\nu^2; m_W, m_l, m_l) \xrightarrow{q^2 \rightarrow 0} C_i (m_N^2, 0, m_\nu^2; m_W, m_l, m_l)\propto m_W^{-2}\,,
\end{equation}
where $i = 0,1,2$  and zero momentum transfer limit $q^{2}=(k_1-p_1)^{2}\rightarrow 0$ (Appendix \ref{appa}) is a good approximation for the evaluation of the loop integral at the low energy regime leading to $C_i\propto m_W^{-2}$. To compute these functions we use the {\tt Package-X} program \cite{Patel:2015tea}.
The effective operator in Eq.~(\ref{effem0}) represents a novel electromagnetic property of normal neutrino and sterile neutrino coupling to photon, 
stemming from the effective right-handed current coupling in Eq.~(\ref{rhc}). 
The same type operator in the quark sector can be obtained, and they 
associate to the effective Dirac mass operator 
$\bar\nu_L N_R$ ($\bar q_L q_R$) of neutrino (quark) 
by the Ward-Takahashi identity \cite{Xue:2016dpl}. This is
different from the effective operator $\bar\nu\sigma_{\mu\nu}\nu F^{\mu\nu}$ of neutrino electromagnetic moment, which has been intensively discussed so far in the literature \cite{Fujikawa:1980yx,Lee:1977tib,Marciano:1977wx,Vogel:1989iv,Broggini:2012df,Giunti:2014ixa}. 
Under the CP transformation of field $A_\mu\rightarrow -A^\mu$ and four-momentum $p^\mu\rightarrow p_\mu$ \cite{Giunti:2014ixa}, it can be shown that 
the effective vertex $\Lambda_{l'}^\mu\rightarrow -\Lambda_{l'\mu}$ and $C_i(q^2)$ are CP invariant. Therefore the effective operator of Eq.~(\ref{effem0}) is CP invariant, except the CP phases 
in the PMNS and $[(U^\ell_R)^\dagger U^\nu_R]$ mixing matrices, 
both are approximately considered as an identity matrix in this article.  

To this end, we have presented a scenario relevant for studying the electron recoil in XENON1T experiment. 
There are three massive sterile neutrinos  $N^e_R$ ($m^e_{N}$), $N^\mu_R$ ($m^\mu_{N})$ and $N^\tau_R$ ($m^\tau_{N}$). They effectively interact with SM charged leptons $e_R$, $\mu_R$ and $\tau_R$, respectively, through the SM gauge bosons $W^{\pm}$ and $Z^{0}$ of weak interaction. Effective EM interaction of sterile neutrinos with SM left-handed neutrinos 
$\nu_L^e$, $\nu_L^\mu$ and $\nu_L^\tau$ is through the SM photon $\gamma$. 
The effective coupling strength ${\mathcal G}_R$ at low energies is a theoretical parameter to be determined experimentally.
In such a scenario,  four-fermion operators of Eq.~(\ref{q1}) induce right-handed current 
couplings \cite{Xue:2016txt,Xue:2016dpl,Xue:2015wha}. We do not need  to add  any fiducial beyond the SM gauge boson, such as $W_R,W',Z'$ or light particles, except right-hand neutrinos. 
In the following, we first constrain the coupling strength ${\mathcal G}_R$ from SM precision measurements,  cosmological and astrophysical observations, then apply this scenario to compute the relevant observable quantities in XENON1T experiment.

\subsection{Sterile Neutrino DM and Coupling Constraints}
It is essential that the contribution of right-handed sterile neutrinos to the total decay width of the W gauge boson  not to exceed the experimental accuracy of W decay width, $4.2\times10^{-2} $ GeV. This gives a weak constraint on the right-handed current coupling as $\mathcal{G}_{R}\lesssim 6\times 10^{-3}$ \cite{Haghighat:2019rht}. A more tighter constraint can be obtained from the radiative decay of massive neutrinos  \cite{Abazajian:2001nj,Gvozdev:1996kx}. It will be shown that the dominate decay channel of massive sterile neutrinos $N_R^l$ is the radiative decay process (Fig. \ref{enu2}) with the corresponding rate as
\begin{align}
\label{decayr}
\Gamma(N_R^l\rightarrow \nu_L^l + \gamma)=
\left( \frac{\alpha \,g_{w}^4}{1024\,\pi^4}\right)m_l^2(m^l_{N})^3\mathcal{G}_R^2 \left[(C_0+2C_1)^2+(C_0+2C_2)(C_0+2C_1) \right]\,.
\end{align}
For different  sterile neutrinos at low energy scale we have
\begin{eqnarray}\label{NRe}
\Gamma(N_R^e\rightarrow \nu_L^e + \gamma)&=&1.57\times 10^{-19}s^{-1} \left(\frac{\mathcal{G}_R}{10^{-4}} \right)^{2}\left(\frac{m_{e}}{511 \,\text{keV}} \right)^{2}\left( \frac{M_{N}^{e}}{100\, \text{keV}}\right)^{3}\,,
\end{eqnarray}
\begin{eqnarray}\label{NRmu}
\Gamma(N_R^\mu\rightarrow \nu_L^\mu + \gamma)&=&6.70\times 10^{-15}s^{-1} \left(\frac{\mathcal{G}_R}{10^{-4}} \right)^{2}\left(\frac{m_{\mu}}{106 \,\text{MeV}} \right)^{2}\left( \frac{M_{N}^{\mu}}{100\, \text{keV}}\right)^{3}\,,
\end{eqnarray}
\begin{eqnarray}\label{NRtau}
\Gamma(N_R^\tau\rightarrow \nu_L^\tau + \gamma)&=&1.87\times 10^{-12}s^{-1} \left(\frac{\mathcal{G}_R}{10^{-4}} \right)^{2}\left(\frac{m_{\tau}}{1.77 \,\text{GeV}} \right)^{2}\left( \frac{M_{N}^{\tau}}{100\, \text{keV}}\right)^{3}\,,
\end{eqnarray}
showing  that decay rates get larger values for  heavier sterile neutrinos and more massive internal leptons. Suppose that  sterile neutrinos $N^e_R$, $N^\mu_R$ and $N^\tau_R$ are DM candidates require that  their lifetimes to be at least $4.4\times 10^{17}$ sec of Universe age. This  yields further constraints on $\mathcal{G}_{R}$ for each type of sterile neutrinos as
\begin{eqnarray}
\mathcal{G}_{R}\lesssim 3.8 \times 10^{-4}\left(\frac{511~ {\rm keV}}{m_e}\right)\left(\frac{100~ {\rm keV}}{m^e_{N}}\right)^{3/2}\,,
\label{gRcon}
\end{eqnarray}
\begin{eqnarray}
\mathcal{G}_{R}\lesssim 1.84 \times 10^{-6}\left(\frac{106~ {\rm MeV}}{m_\mu}\right)\left(\frac{100~ {\rm keV}}{m^\mu_{N}}\right)^{3/2}\,,
\label{gRconmu}
\end{eqnarray}
\begin{eqnarray}
\mathcal{G}_{R}\lesssim 1.1 \times 10^{-7}\left(\frac{1.77~ {\rm GeV}}{m_\tau}\right)\left(\frac{100~ {\rm keV}}{m^\tau_{N}}\right)^{3/2}\,.
\label{gRcontau}
\end{eqnarray}
In the seesaw mechanism, there is a transition between active SM neutrinos and sterile neutrinos which is induced by a "mixing angle". This mixing would impact on the relic abundance of sterile neutrinos indirectly through the interaction of generated active neutrinos with thermal bath of SM particles \cite{PhysRevLett.72.17,Gelmini:2004ah,Abazajian:2001nj}. However in the model  presented here, there is an effective interaction between right-handed sterile neutrinos  and SM gauge bosons (\ref{rhc}) leading to the effective scattering of sterile neutrinos from SM particles (Fig.~\ref{enu2}). These  interactions contain the effective coupling to right-handed currents (see diagrams in Fig.~\ref{fig1}), and can be compared to those effective interactions in seesaw model, which are induced by the mixing angle and follows by a conversion process between active and sterile neutrinos. If one replaces the right-handed coupling $\mathcal{G}_{R}$ in Eq.(\ref{decayr}) by active-sterile mixing angle $\theta_{N\nu}$, the only deviation from the radiative decay rate obtained in  the context of seesaw model \cite{Lee:1977tib,PhysRevD.25.766,Adhikari:2016bei,Abazajian:2001nj}, is the appearance of $m_{\ell}^{2}M_{N}^{3}$ instead of $M_{N}^{5}$. This specific property results from the effective couplings (\ref{rhc}) and (\ref{effem0}) of right-handed sterile neutrinos to SM sector. 
Although in seesaw model the radiative decay channel  $N\rightarrow \nu +\gamma$ is a sub-dominant process and the lifetime of sterile neutrino DM is determined by  $N\rightarrow \nu_l \bar \nu_{\alpha} \nu_{\alpha}$ \cite{Barger:1995ty,Abazajian:2001nj}, the latter is absent in the context of our model and the radiative decay of sterile neutrinos turns out to be the dominant decay channel.

 There are various theoretical frames explaining   sterile neutrino with keV-scale masses which can provide the relic density of DM \cite{Asaka:2005an,Mohapatra:1986bd}. We estimate the relic density  of sterile neutrinos by using the interaction rate and the Hubble rate during the radiation dominated era  \cite{Olive:2003iq,Hansen:2001zv,Lin:2019uvt,Kolb:1990vq,Biswas:2018iny}
\begin{align}\label{2.7e}
\Gamma_{N}=n_{N}\langle \sigma v \rangle \approx G_{F}^{2}\mathcal{G}_{R}^{2} T^{5}\,, \,\,\,\,\,
H=\sqrt{\frac{4\pi^3 g_{\rho}  }{45}}
\frac{T^{2}}{M_{Pl}}\approx1.66 g_{\rho}^{1/2} \frac{T^{2}}{M_{Pl}}
\,,
\end{align}
where the energy density degrees of freedom is shown by $g_{\rho}$ \cite{Drees:2015exa} and $n_N$  denotes the sterile neutrino number density. When $\Gamma_{N}\approx H$, sterile neutrinos decouple at temperature 
\begin{align}
T_{N}\approx 10 \ \text{GeV} \left(\frac{10^{-6}}{\mathcal{G}_{R}}\right)^{2/3}g_{\rho}^{1/6},
\label{TNdec}
\end{align}
note that  SM neutrinos decoupled at 
$T_{\nu}\approx 1 ~\text{MeV}$. Since the interaction of  sterile neutrinos is weaker than SM neutrinos, they decoupled earlier than  SM types in the early Universe. 
The number density of sterile neutrinos is obtained as
\begin{align}
\rho_{N}=m_{N} n_{N}= m_{N} Y_{N} n_{\gamma}\,,
\end{align}
where $Y_{N}=n_{N}/n_{\gamma}$ is the fraction of density of sterile neutrinos relative to the density of photons. Therefore the relic abundance of keV neutrinos produced during the radiation dominated epoch can be found as \cite{Biswas:2018iny}
\begin{align}\label{relic}
\Omega_{N}h^{2}\approx  
76.4 \left[ \frac{3g_{N}}{4g_{ s}(T_{N})}\right] \left(\frac{m_{N}}{\text{keV}}\right)\,,
\end{align}
which $g_{N}$ is the sterile neutrino degrees of freedom  and $g_{s}$ denotes the entropy density number of relativistic degrees of freedom of thermal bath particles \cite{Drees:2015exa}. 
DM relic density presented in Eq.~(\ref{relic}) obviously demands entropy dilution process after freeze-out, to bring down the abundance consistent with the current DM relic density obtained from CMB observations  $\Omega_{DM}h^{2}=0.120\pm 0.001$ \cite{Aghanim:2018eyx}. If the coupling limit does not match with the DM produced in the radiation dominated era, a possible matter dominated or kination scenario can explain the mismatch of the relic abundance \cite{Allahverdi:2020bys,Boyarsky:2018tvu}.  Besides, there is another possibility which can be considered. Then we might consider   
 the oscillation of keV sterile neutrino  to heavier sterile neutrinos after its decoupling \cite{Xue:2016dpl},  analogous to the SM neutrino oscillation.  Sterile neutrinos  decay to the SM particles, leading to keV DM relic abundance observed today. Moreover, considering sterile neutrinos may influence on the extra  number of effective neutrinos $\Delta N_{\text{eff}}$ decoupled from the thermal bath in the early Universe \cite{Drewes:2013gca}. The parameter $\Delta N_{\text{eff}}$ is constrained by CMB \cite{Aghanim:2018eyx}. Its exact value in our scenario also depends on the cosmic history before the big bang nucleosynthesis and after the inflationary era~\cite{Gelmini:2019esj,Drewes:2013gca} which is unknown then we will not consider its computation here. 

%%%%%%%%%%%%%%%%%%%%%%%%%%%%%%%%%%%%%%%%%%%%%%%%%%%%%%%%%%%%%%%%%%%%%%%%%%%%%%%%%%%%%%%%%%%%%%%%%%%%%%%%%%%%%%%%%%%%%%%%%%%%
%%%%%%%%%%%%%%%%%%%%%%%%%%%%%%%%%%%%%%%%%%%%%%%%%%%%%%%%%%%%%%%%%%%%%%%%%%%%%%%%%%%%%%%%%%%%%%%%%%%%%%%%%%%%%%%%%%%%%%%%%%%%
%%%%%%%%%%%%%%%%%%%%%%%%%%%%%%%%%%%%%%%%%%%%%%%%%%%%%%%%%%%%%%%%%%%%%%%%%%%%%%%%%%%%%%%%%%%%%%%%%%%%%%%%%%%%%%%%%%%%%%%%%%%%
%
\section{Neutrino-Electron Scattering}
\label{sec:scat}
\subsection{SM Neutrino-Electron Scattering}

In the low energy regime of the SM,  when the momentum carried by intermediate vector bosons $W^{\pm}$ and $Z^{0}$ are much less than their masses, the effective Lagrangian
describing the neutrino-electron scattering
via the charged current (CC) or neutral current (NC)   are given by
\begin{align}
\mathcal{L}^{CC}_{\text{eff}}=-2\sqrt{2} G_{F} \Big([\bar \nu_{e} \gamma^{\mu} P_{L} u_e][\bar u_e \gamma_{\mu} P_{L} \nu_{e}]   \Big)  \,,
\end{align}

\begin{align}
\mathcal{L}^{NC}_{\text{eff}}=-\sqrt{2} G_{F} \sum_{\alpha=e,\mu\tau}
\Big(\bar u_e \gamma_{\beta} (g^{\alpha}_{L}P_{L}+g^{\alpha}_{R}P_{R})u_e\Big)[\bar \nu_{\alpha} \gamma^{\beta} P_{L} \nu_{\alpha}]\,,
\end{align}
where $P_{R,L}=\frac{1}{2}(1\pm\gamma^{5})$, and $g^{\alpha}_{R}=2\sin^{2}\theta_{W}\pm1$ with $"+"$  for electron neutrino flavour $\alpha=e$, and $"-"$ for $\alpha=\tau,\mu$ and $g^{\alpha}_{L}=2\sin^{2}\theta_{W}$ for all the neutrino flavours $\alpha=e,\mu,\tau$. Note that $\sin^{2}\theta_{W}=0.23$ represents the weak mixing angle. The corresponding cross section of the neutrino-electron scattering induced by these low-energy current interactions is given by \cite{Mohapatra:1998rq,Radel:1993sw,Formaggio:2013kya,Tomalak:2019ibg}
\begin{align}
\frac{d\sigma^{e\nu_{\alpha}\rightarrow e\nu_{\alpha}}_{SM}}{dE_{r}}=\frac{2G^{2}_{F}m_{e}}{\pi} \left[
(g^{\alpha}_{L})^{2}+(g^{\alpha}_{R})^{2} \left(1-\frac{E_{r}}{E_{\nu}}\right)^{2}-g^{\alpha}_{R}g^{\alpha}_{L}\frac{m_{e}E_{r}}{E_{\nu}^{2}} \right]\,, 
\end{align}
where $E_{r}$ is the recoil energy of electron in the detector and $E_{\nu}$ is the energy of incoming neutrinos\footnote{Note that this cross section can be also re-written in terms of  $g_{V}$ and $g_{A}$ where $g_{V,A}=\frac{1}{2}(g_{R}\pm g_{L})$ as  
\begin{align}
\frac{d\sigma_{SM}}{dE_{r}}=\frac{G^{2}_{F}m_{e}}{2\pi} \left[
(g_{V}+g_{A})^{2}+(g_{V}-g_{A})^{2} (1-\frac{E_{r}}{E_{\nu}})^{2}+(g_{A}^{2}-g_{V}^{2})^{2}\frac{m_{e}E_{r}}{E_{\nu}^{2}} \right]\,.
\end{align}}.

\subsection{Sterile Neutrino-Electron  Scatterings}
We compute the contribution of sterile neutrinos to the electron recoil event.  
In the model presented in Sec.~\ref{ch2},
sterile neutrino  interactions with electrons  give rise to additional contributions to the recoil process. The effective Lagrangian describing the low energy scattering between sterile neutrinos and electrons regarding  effective vertices presented in Fig.~\ref{fig1}, are given by

\begin{eqnarray}\label{Scc}
\mathcal{L}^{SCC}_{\text{eff}}&=&-2\sqrt{2} G_{F} \Big([\bar N \gamma^{\mu} \mathcal{G}_{R} P_{R} u_e][\bar u_e \gamma_{\mu} P_{L} \nu_{e}] +[\bar N \gamma^{\mu} \mathcal{G}_{R} P_{R} u_e][\bar u_e \gamma_{\mu} \mathcal{G}_{R} P_{R} N]  \Big)\\ \nonumber &=&-2\sqrt{2} G_{F} \Big([\bar N \gamma^{\mu}\mathcal{G}_{R} P_{R} u_e][\bar u_e \gamma_{\mu} P_{L} \nu_{e}] \Big)+G_{F}\mathcal{O}(\mathcal{G}_{R}^{2})\,
\end{eqnarray}
and 
\begin{align}\label{SNC}
\mathcal{L}^{SNC}_{\text{eff}}=-\sqrt{2} G_{F} 
\Big(\bar u_e \gamma_{\beta} (g_{L}P_{L}+g_{R}P_{R})u_e\Big)[\mathcal{F}_{R}\bar N \gamma^{\beta} P_{R} N]\,.
\end{align}
In principle there are two possibilities in which sterile neutrinos can participate in the electron recoil event; 
\begin{enumerate}[I)]
  \item Sterile neutrinos may appear as incoming particle contributing to the galactic DM halo where initial flux of sterile neutrinos can be inferred from the local density of DM, in this case our final states are recoiled electrons and SM neutrinos.
  \item Sterile neutrinos may appear as outgoing particle whether or not dark matter is made of sterile neutrinos, in this case initial flux is provided by solar neutrinos leading to emission of sterile neutrinos after recoil events.
\end{enumerate}
In general, scattering processes can  be divided into two types; elastic and inelastic. It is crucially important to clarify, in which types of scattering processes and also scattering channels,  sterile neutrinos  can produce recoil signal in the  desired energy range detected by XENON1T experiment. It has been recently shown that an elastic scattering process between a  DM particle and electron, only if DM satisfies velocity  $v_{DM}\sim 0.1$ and mass $m_{DM}\gtrsim 0.1$ MeV,   can produce the electron recoil excess in the energy range  of  $\mathcal{O}(keV)$  \cite{He:2020wjs,Kannike:2020agf,Alhazmi:2020fju}. While such a high velocity  is far above   the local escape velocity of  the Milky Way $\sim 10^{-3}$, it is argued that a long range attractive force might accelerate DM towards the Earth and boosted to high velocities near the Earth surface \cite{Davoudiasl:2020ypv}.
 In contrast to the elastic scattering where the kinetic energy of a non-relativistic light DM is not sufficient to produce  recoil electron at keV scale, an inelastic DM-electron scattering thanks to the momentum transfer in the order of $\sim m_{DM}$ leads to the electron kinetic energy  $T_{e}\approx m_{DM}^{2}/2m_{e} =2.45 \ \text{keV} (m_{DM}/50~ \text{keV})^{2}$ at the final state  \cite{Kannike:2020agf,Harigaya:2020ckz,Campos:2016gjh,Baryakhtar:2020rwy,He:2020wjs,Ando:2010ye}. 
 
The DM inelastic scattering from atomic electrons may lead to electron ionization or electronic excitation, and DM-nuclear scattering may cause molecular dissociation \cite{Essig:2011nj,Liao:2013jwa,Gounaris:2004ji}.  For electron ionization, the total energy $E_{N}$ of an incoming sterile neutrino must be larger than the energy of a bound electron  in level $i$, this defines  a threshold condition
for the process to occur in the various bound states (see Eq. \ref{dsigmatotal}).  It has been shown that the  scattering from bound electrons leads to larger recoil than the free electron case \cite{Campos:2016gjh,Liao:2013jwa}. 
 In fact the effect of bound electrons in atomic shells is important at low-energy electronic recoil around a few keV and  causes some deviations from the approximation of free electron \cite{Chen:2016eab,Gounaris:2004ji,Chen:2013lba,Roberts:2019chv,Liao:2013jwa,Chen:2014ypv}. 
 
In the following we consider inelastic scattering of the sterile neutrinos with electrons taking into account effective interactions given by Eqs. (\ref{Scc})  and (\ref{SNC}).

\subsubsection{$N_{e}e\rightarrow \nu_{e}  e$ }\label{s321}
The differential cross section of $Ne\rightarrow \nu_{e} e$  corresponding to the first diagram  in Fig. (\ref{Fig22}), is obtained as follows
\begin{eqnarray}
\label{neenu}
\frac{d\sigma(N_e e\rightarrow e\nu_{e})}{dE_{r}}&=&
\frac{2G_{F}^{2}}{\pi}  \frac{\mathcal{G}_{R}^{2}m_{e}}{|\vec p_{N}|^{2}}E_{N}\left[E_{N}+\frac{m_{N}^{2}}{2m_{e}}\right]\,, 
\end{eqnarray}
where $|\vec p_{N}|=(E_{N}^{2}-m_{N}^{2})^{1/2}$ is the momentum of  sterile neutrinos. We assume that the sterile neutrinos as DM have the same  velocity as in the standard halo model  $v_{N}\approx 220 \text{km/s}$ and $E_{N}\sim m_{N}$, for  sterile neutrinos with masses 
in $\mathcal{O}(\text{keV})$ range 
\begin{eqnarray}
\frac{d\sigma(N_e e\rightarrow e\nu_{e})}{dE_{r}}\approx
1.94 \times 10^{-53} \left( \frac{\mathcal{G}_{R}}{10^{-6}}\right)^{2} \left(\frac{10^{-3}}{v_{N}} \right)^{2} \text{cm}^{2}\left[\text{keV}^{-1}\right]
\,, 
\end{eqnarray}
obviously right hand side of Eq. (\ref{neenu})  is independent of the recoil energy $E_{r}$ and  only provide  a  flat recoil spectrum which  can not be used to explain XENON1T excess at low energy. 

\subsubsection{$N_{\ell}e\rightarrow N_{\ell} e$ }\label{s322}

\begin{figure}
  \centering
    \subfloat{
    \centering
    \includegraphics[width=1.5in]{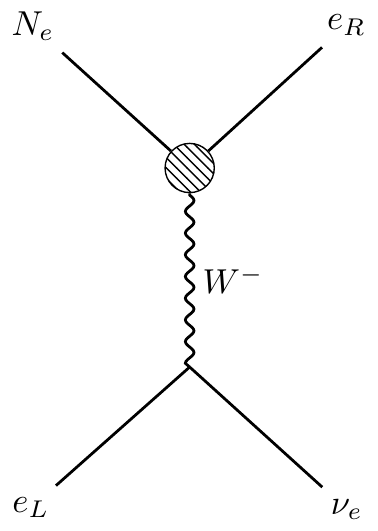} 
    \hspace{2cm}
    \includegraphics[width=1.5in]{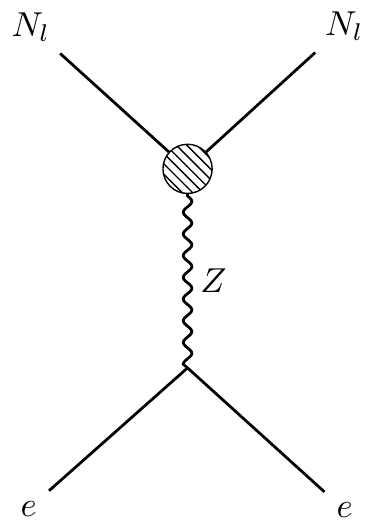}
    }
\caption{ Feynman diagrams corresponding to sterile neutrino-electron scattering through both the charged current and neutral current. In these diagrams time flows from left to right. 
}\label{Fig22}
\end{figure}

If we consider the elastic scattering between  sterile neutrinos and electrons (second diagram  in Fig. (\ref{Fig22})), the differential cross section is given by 
\begin{equation}
\label{nene}
\frac{d\sigma(N_\ell e\rightarrow N_\ell e)}{dE_{r}}= \frac{G_{F}^{2}}{2\pi}  \frac{\mathcal{G}_{R}^{2}m_{e}}{|\vec p_{N}|^{2}} \left[
(g^{e}_{L})^{2}(E_{N}-E_{r})^{2}+(g^{e}_{R})^{2} E_{N}^{2}
-g^{e}_{L}g^{e}_{R}(m_{N}^{2}-m_{e}E_{r})
\right]\,.
\end{equation}
Since in such an elastic scattering, there is not enough momentum transfer  provided by non-relativistic keV mass sterile neutrinos, it is not possible to produce observable signal from the electron recoil in $N\,e\rightarrow N\,e$ process. Note that an extra contribution to $N\,e\rightarrow N\,e$ originates from W exchange which is at higher order $\mathcal{O}(\mathcal{G}_R^4$) and we ignore it here.

\subsubsection{$N_\ell e\rightarrow \nu_{\ell} e$ 
 (EM Channel)}\label{s323}
 \label{loopincsec}
Here we consider electron-neutrino scattering via electromagnetic (EM) channel \cite{Giunti:2014ixa,Broggini:2012df}. % 
The neutrino magnetic-moment $\mu_{\nu}$ contribution to the electron-neutrino scattering process is given by \cite{Vogel:1989iv}
\begin{eqnarray}
\frac{d\sigma{(e\nu_{e}\rightarrow e\nu_{e}})}{dE_r}=\frac{\pi \alpha^{2} }{m_{e}^{2}} \left( \frac{1}{E_{r}}-\frac{1}{E_{\nu}} \right) \left( \frac{\mu_{\nu}}{\mu_{B}} \right)^{2}\,,
\end{eqnarray}
where $\mu_{B}$ is the Bohr Magneton. 
For neutrino-electron scattering the magnetic moment contribution is dominant over the standard electroweak contribution at low recoil energies. However, a strict  astrophysical bound on the neutrino magnetic moment is provided  by the observation of  properties of globular cluster stars and stellar cooling process  $\mu_{\nu}\lesssim 10^{-12}\mu_{B}$ \cite{Tanabashi:2018oca,2019arXiv191010568A,Raffelt:1996wa}.  The upper bound for the neutrino magnetic moment gives $d\sigma/dE_{r}\approx 10^{-45} \text{cm}^{2}\text{keV}^{-1}(\text{keV}/E_{r})$, it is claimed in  Ref.~\cite{Aprile:2020tmw} that this $1/E_{r}$ enhancement at low energy could provide an explanation for XENON1T excess but with less statistical significance compared to other alternative possibilities. 

\begin{figure}
    \centering
    \includegraphics[width=2in]{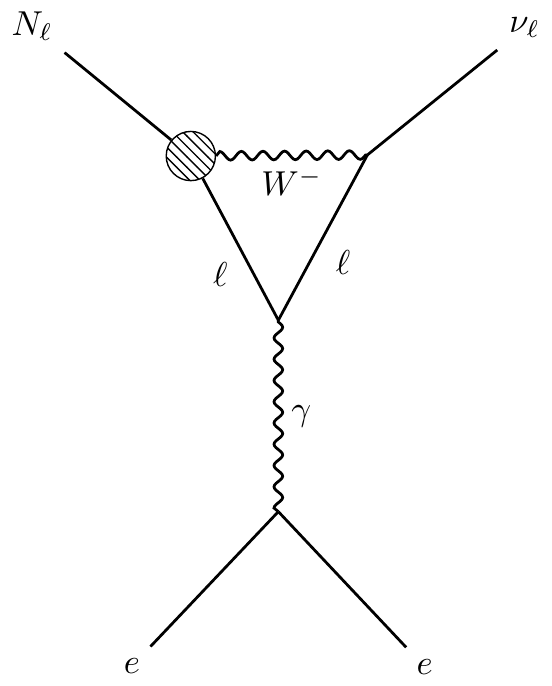}
    \caption{ Sterile neutrino-electron  inelastic (EM channel) scattering.}
    \label{enu}
\end{figure}

In the following, we consider the sterile neutrino EM scattering from electrons through lepton loop correction (see Fig.~\ref{enu}), namely through the effective EM vertex of Eq.~(\ref{effem0}) or Fig.~\ref{enu2}. To find the cross section of  $Ne\rightarrow \nu_{l} e$ regarding leptons $l=e, \mu, \tau$ at loop level of the same family of SM and sterile neutrinos,  we use Eq.~(\ref{crossincom}) and the kinematics used for its derivation. Then the  differential cross section is given by 
\begin{align}\label{crossloop}
    \frac{d\sigma{(N_\ell e\rightarrow \nu_{l} e})}{dE_{r}}=& ~\left(\dfrac{m_l^2}{8\pi\, m_{e} \, q^4\,p_{N}^{2}} \right) \, \left( \dfrac{\alpha \, g_w^2 \, \mathcal{G}_R }{4 \, \pi} \right)^2  (2m_N m_\nu +2m_{e}E_{r}+ M_{N}^{2}) \times  \\ \nonumber
    &\big\{ (2C_1+C_0)^2 \, [m_e(2m_e E_N+M_N^{2})(E_N-E_r))] + 
    \\ \nonumber
    &\;\;(2C_2+C_0)^2 \, [m_e(2M_N E_r +M_N^2)(E_N-E_r)]+  \\
    \nonumber
    &\;\;(2C_1+C_0)(2C_2+C_0)\;\;[2m_e^2 (E_N^2+E_r^2-2E_N E_r)+m_e m_N^2 (E_N-E_r) ]\big\}\,. \nonumber
\end{align}
 The recoil energy transfers to an electron through the inelastic scattering of sterile neutrino  $q^2=(k_1-p_1)^2=2 m_e E_r$ (Appendix \ref{appa}). Here we assume vanishing neutrino mass $m_{\nu}\approx 0$ in favour of the  sterile neutrino and electron masses.  To give more insight at the low energy scale, Eq. (\ref{crossloop}) can be approximated as
\begin{eqnarray}
\frac{d\sigma_{N_\ell e\rightarrow \nu_{l} e}}{dE_{r}}&\propto&  
\!\left(\frac{\alpha G_{F} \mathcal{G}_{R}m_{\ell}}{4 \pi^{3/2}v_{N}E_{r}m_{e}^{1/2}} \right)^{2}
(M_{N}^{2}+2m_{e}E_{r})\,.
\end{eqnarray}

\subsubsection{ $\nu_{e}e\rightarrow N_e e$}\label{s324}
The cross section of neutrino-electron scattering when a sterile neutrino contribution as an outgoing particle is 
\begin{equation}
\label{outgocross}
\frac{d\sigma(\nu_{e} e\rightarrow N_e e)}{dE_{r}}=\left(\frac{G_{F}^{2}\mathcal{G}_{R}^{2}|\vec p_{N}|^{2}}{2\pi E_{\nu}E_{N}^{2}}\right)\frac{\left[m_{N}^{2}+2m_{e}E_{r} \right](2m_{e}E_{\nu}-m_{N}^{2})}{m_{e}(E_{r}+|\vec p_{N}|)+E_{\nu}(|\vec p_{N}|-E_{N})+m_{N}^{2}}\,,
\end{equation}
here $E_{N}=E_{\nu}-E_{r}$ is the energy of outgoing sterile neutrinos in terms of the energy of standard model neutrinos $E_{\nu}$ and the recoil energy of electrons $E_{r}$, and $|\vec p_{N}|=\sqrt{E_{N}^{2}-m_{N}^{2}}$. Our estimation shows this process will not provide a significant contribution to the XENON1T excess (see Appendix \ref{appest} for the outgoing process).
We will not compute the cross section for the process $\nu_{\ell} e\rightarrow N_\ell e $ at one loop level like Sec.~\ref{loopincsec} when the mediator is photon. Since it will not give a large enough cross section to explain the excess.
%

%%%%%%%%%%%%%%%%%%%%%%%%%%%%%%%%%%%%%%%%%%%%%%%%%%%%%%%%%%%%%%%%%%%%%%%%%%%%%%%%%%%%%%%%%%%%%%%%%%%%%%%%%%%%%%%%%%%%%%%%%%%%
%%%%%%%%%%%%%%%%%%%%%%%%%%%%%%%%%%%%%%%%%%%%%%%%%%%%%%%%%%%%%%%%%%%%%%%%%%%%%%%%%%%%%%%%%%%%%%%%%%%%%%%%%%%%%%%%%%%%%%%%%%%%
%%%%%%%%%%%%%%%%%%%%%%%%%%%%%%%%%%%%%%%%%%%%%%%%%%%%%%%%%%%%%%%%%%%%%%%%%%%%%%%%%%%%%%%%%%%%%%%%%%%%%%%%%%%%%%%%%%%%%%%%%%%%
%

\section{Calculation the Electronic Recoil Events}
\label{sec:recoil}
We consider two different possibilities for the sterile neutrino interacting with electrons in the XENON detector which might explain the reported anomaly. 

\subsection{Incoming Sterile Neutrino}
The differential number of events in reconstructed electron recoil energy $E_{rec}$, for the inelastic scattering of sterile neutrino DM with electrons per unit detector mass as observed by XENON1T is 
%,
\begin{eqnarray}
\label{ingevent}
\frac{dR}{dE_{rec}}= N_T \, \frac{\rho_{DM}}{m_{N}} \, \int dE_r \int_{v_{\text{min}}}^{v_{\text{esc}}} dv \, v \, f(v)   
\, \frac{d\sigma_{tot}}{dE_r} \, \epsilon (E_r) \, G(E_{rec},E_r)\,.
\end{eqnarray}
Here $\rho_{DM}$ is the local DM density, $m_{N}$ is the mass of the sterile neutrino, $f(v)$ is the DM halo velocity distribution in the lab frame, $\epsilon(E_r)$ is the detector response efficiency and $d\sigma_{tot}/dE_{r}$ is the differential cross section (see Appendix \ref{appest} for the estimation of event rate). We use a step-function approximation for the energy of bound electrons in XENON atoms  \cite{Aprile:2020tmw,Chen:2016eab,Chen:2013lba} 
\begin{eqnarray}
\frac{d\sigma_{tot}}{dE_r}=\sum^{54}_{i=1} \Theta (E_r-B_i) \frac{d\sigma}{dE_r}\,,
\label{dsigmatotal}
\end{eqnarray}
where the binding energy for an electron in shell $i$ is $B_i$ using the data of Ref.~\cite{NIST_ASD}. In fact, one would expect a sharp enhancement in the differential cross section when reaching to a new atomic shell, the stepping approximation in Eq. (\ref{dsigmatotal}) is  used to set an upper bound here \cite{Chen:2016eab,Chen:2013lba}. This approximation is consistent with more robust calculations involving the internal structure of the XENON nucleus \cite{Hsieh:2019hug}.  Meanwhile, we use the data from Ref.~\cite{Aprile:2020tmw} for the detector efficiency $\epsilon(E_r)$ as a function of recoil energy $E_r$. The number of events are then smeared with a Gaussian distribution $G(E_{rec},E_r)$ with $\sigma(E_r)/E_r  = (0.3171/\sqrt{E_r [keV]}) + 0.0015 $ to account for the response of the photo-multiplier tubes %(PMTs)
as a function of the reconstructed recoil energy in the XENON1T detector \cite{Aprile:2020yad}.
For integrating the halo velocity distribution we use closed-form analytic expressions~\cite{Drees:2019qzi,Barger:2010gv,DelNobile:2013sia} and assume a  standard 
DM  halo with $\rho_{DM} \simeq  0.4 \text{ GeV cm}^{-3}$ \cite{Weber:2009pt}, the local circular velocity $v_0= 220$ km/s, the most probable velocity $v_e= 232$ km/s and the escape velocity $v_{\text{esc}} = 544$ km/s. The minimum energy that includes the relevant minimum velocity is described in Appendix \ref{appa}.

\subsection*{Fitting with the XENON1T Excess}
We fit the XENON1T data with our model for the incoming sterile neutrino DM scatters  from electrons to the SM neutrinos using Eq.~(\ref{ingevent}) to obtain the required number of excess events. Moreover, we consider the $\chi^2$ test to check the reliability of our fit. As a consequence,  we use \cite{Miranda:2020kwy,Davoudiasl:2020ypv,Dessert:2020vxy}
\begin{eqnarray}
\label{eq:chi2fit}
\chi^2=\sum_j 
\frac{1}{\sigma_j^2}\left[\frac{dR_{exp}}{dE_{rec}}-\frac{dR_{th}}{dE_{rec}}\right]^2\,,
\end{eqnarray}
where $\sigma_i$ is the variance of experimental data by XENON1T.  The statistical significance of our fit can be identified by $\delta \chi^2=\chi_{min}^2 - \chi_{bg}^2$ where $\chi_{bg}^2\simeq46.5$ for the background estimated by the XENON Collaboration \cite{Davoudiasl:2020ypv,Dessert:2020vxy}. The parameter $\chi_{min}^2$ is the minimum value of $\chi^2$ in Eq.~(\ref{eq:chi2fit}) that can be found by scanning over  a range of values for the coupling $\mathcal{G}_R$ and the sterile neutrino mass $m^\ell_N$. The statistical importance of background is shown by $\chi_{bg}^2$ calculated from the background estimation and the measured data points by the XENON1T experiment.
To match the presence of sterile neutrino DM with the data of XENON1T excess one requires  $\delta \chi \simeq 3.2$ which is reported by the experiment  \cite{Aprile:2020tmw}.
The scattering processes of  incoming sterile neutrinos given by the cross sections in Eqs.~(\ref{neenu}) and (\ref{nene}), do not have $1/E_{r}$ enhancement factor to reproduce event excess at low energy recoil spectrum.  However, they may have a spectrum at higher energies assuming a large enough coupling that is constrained by DM lifetime in Eq.~(\ref{gRcon}). Then, in practice we can not have any observational effect from these processes in XENON1T experiment.

However, the sterile neutrino loop interaction  described in Fig.~\ref{enu} and computed in Eq.~(\ref{crossloop}), besides giving enhancement recoil energy factors at low energy, due to the presence of the internal loop leptons masses enhance the cross section. Consequently, it can provide the  sufficient event rate for specific choices of coupling and mass.

\begin{figure}
  \centering
    \subfloat{
    \centering
    \includegraphics[width=2.4in]{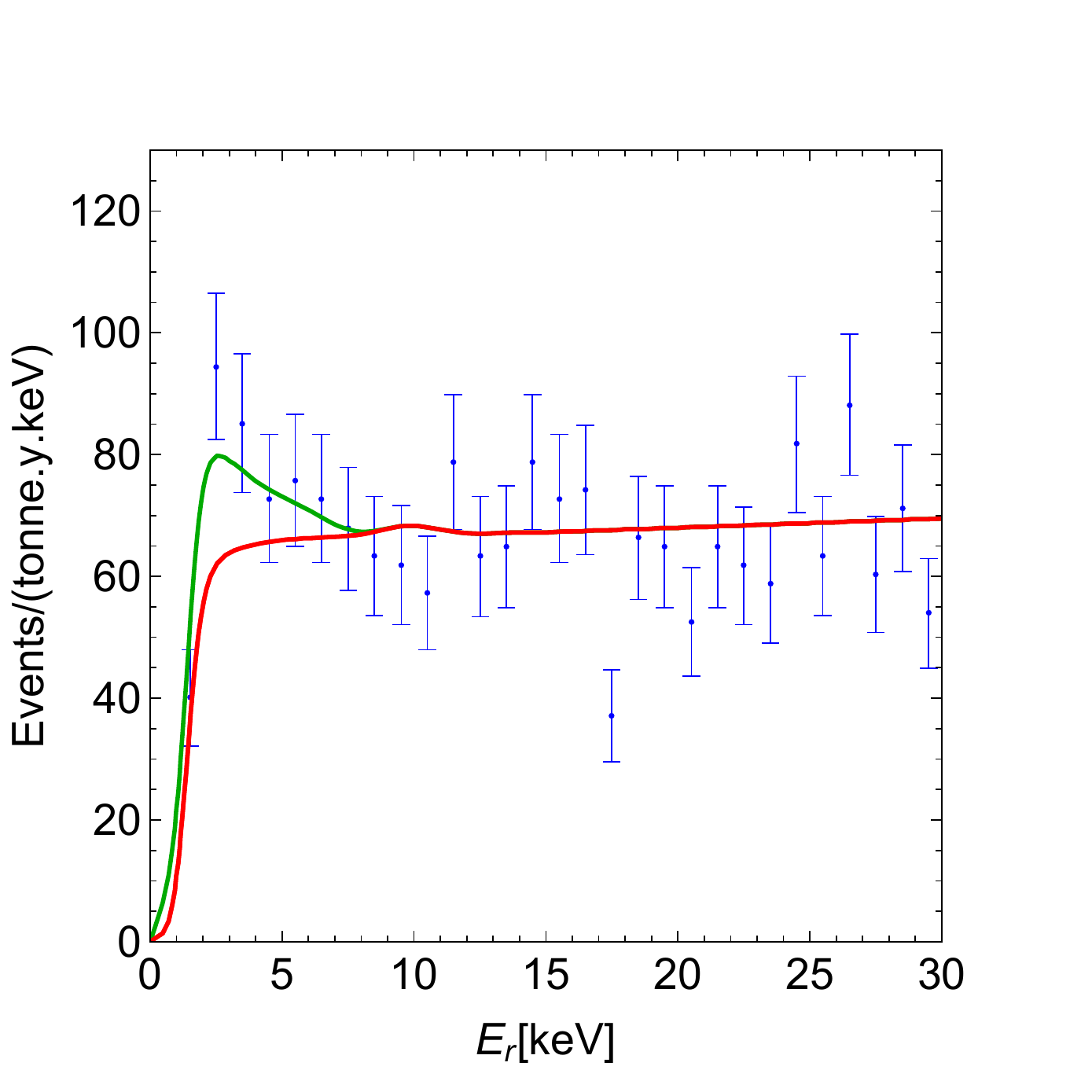} (a) \includegraphics[width=2.4in]{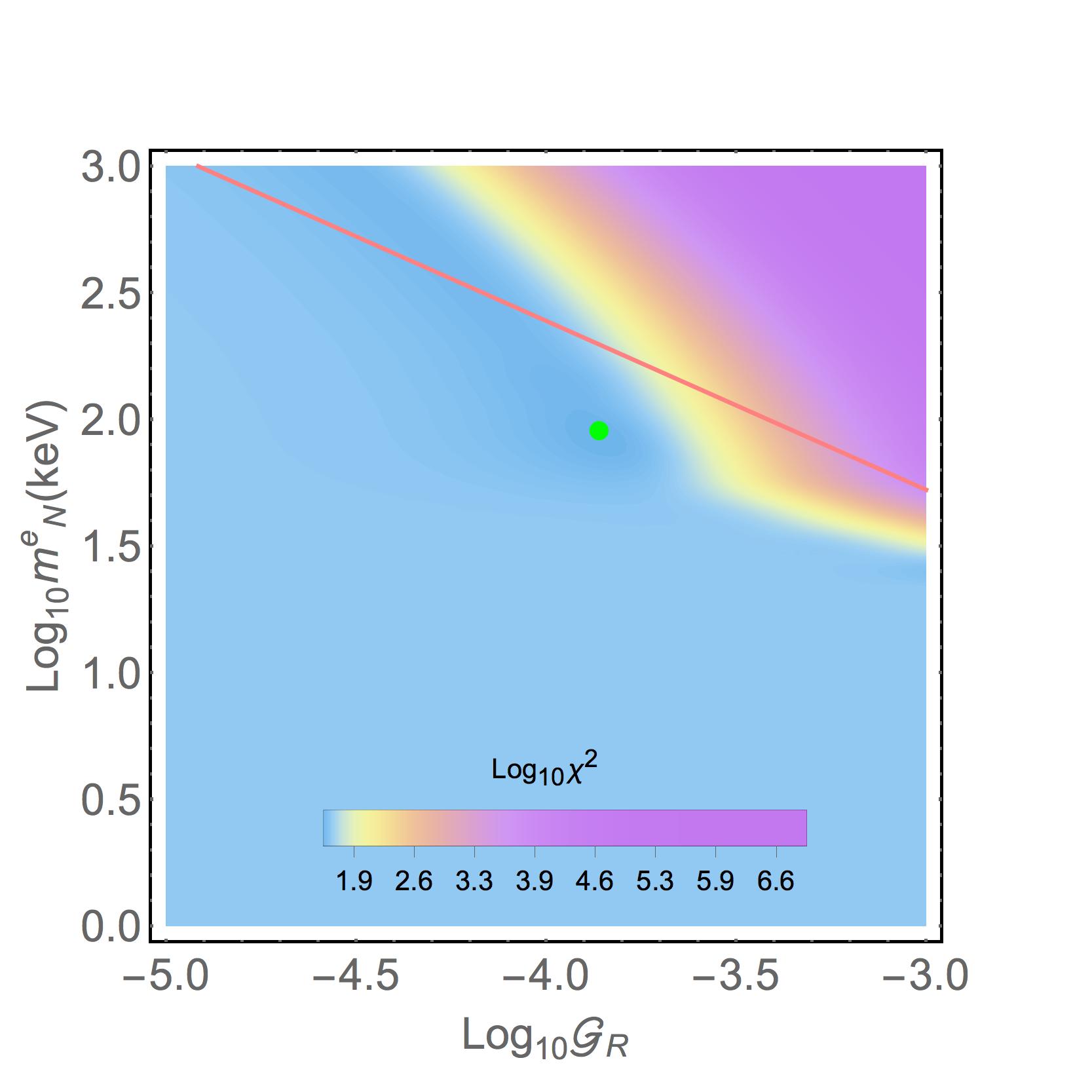}
    \label{fig:event-electron}
    }
    \\
    \subfloat{
    \centering
    \includegraphics[width=2.4in]{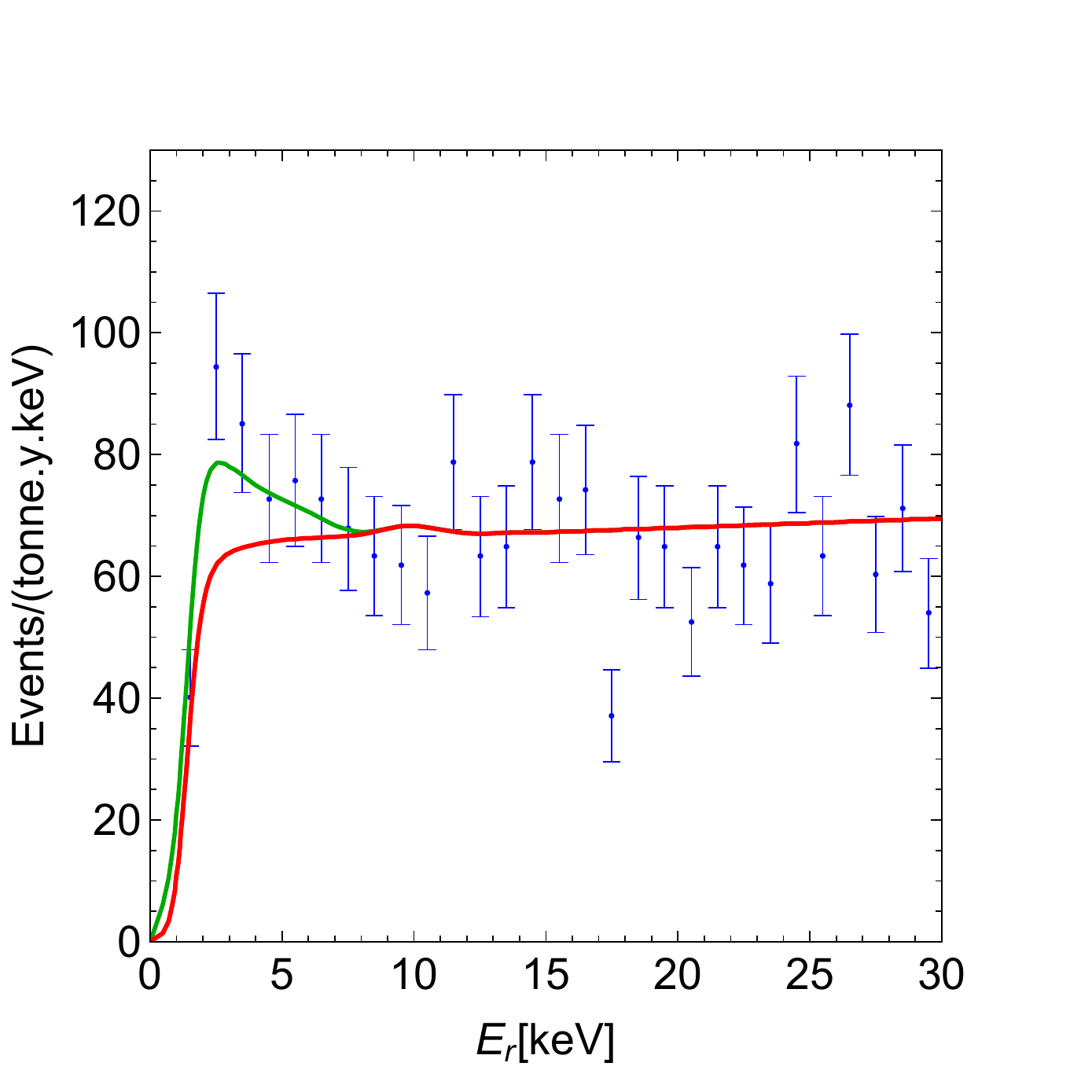}  (b) 
    \includegraphics[width=2.4in]{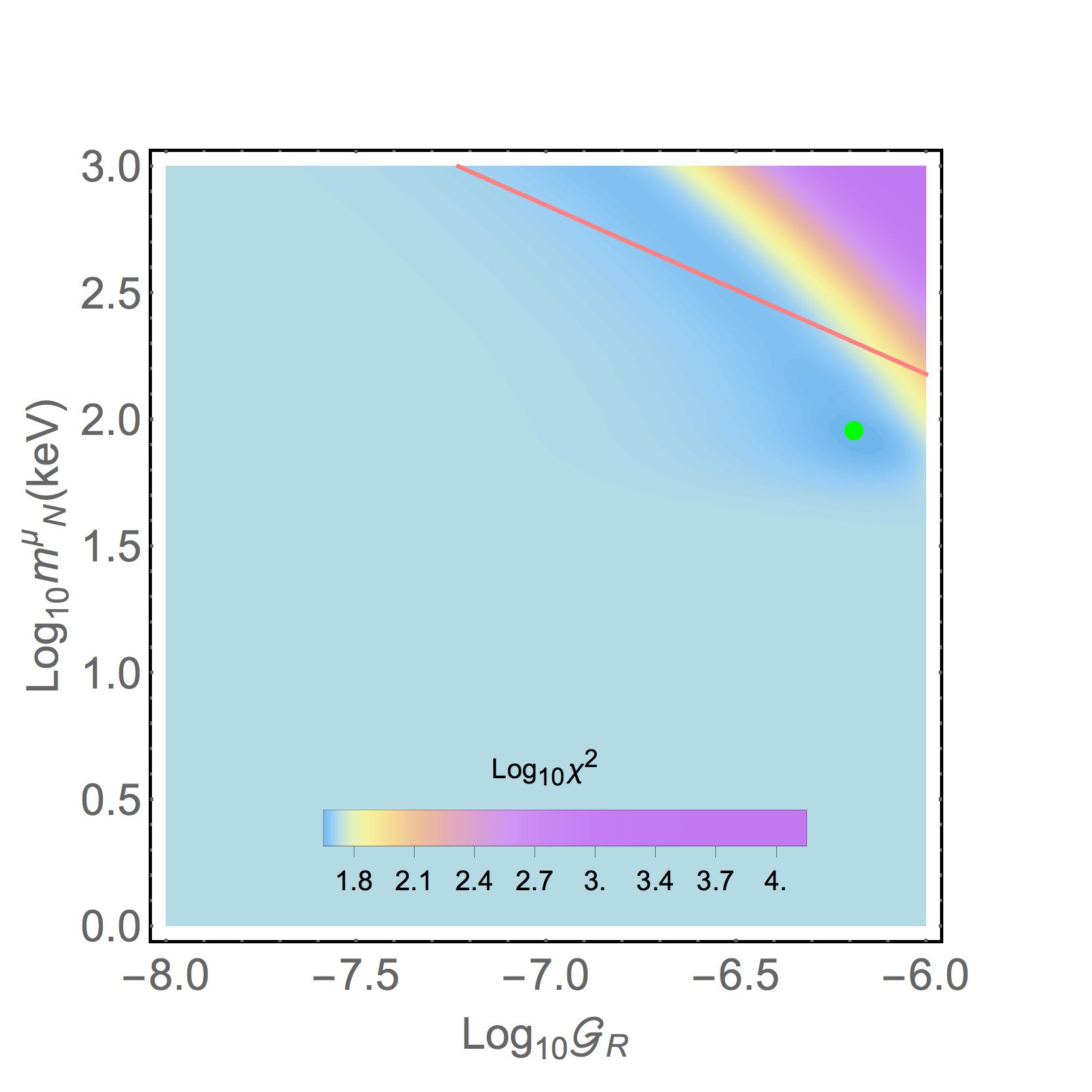}
    \label{fig:event-muon}
    }
    \\
    \subfloat{
    \centering
    \includegraphics[width=2.4in]{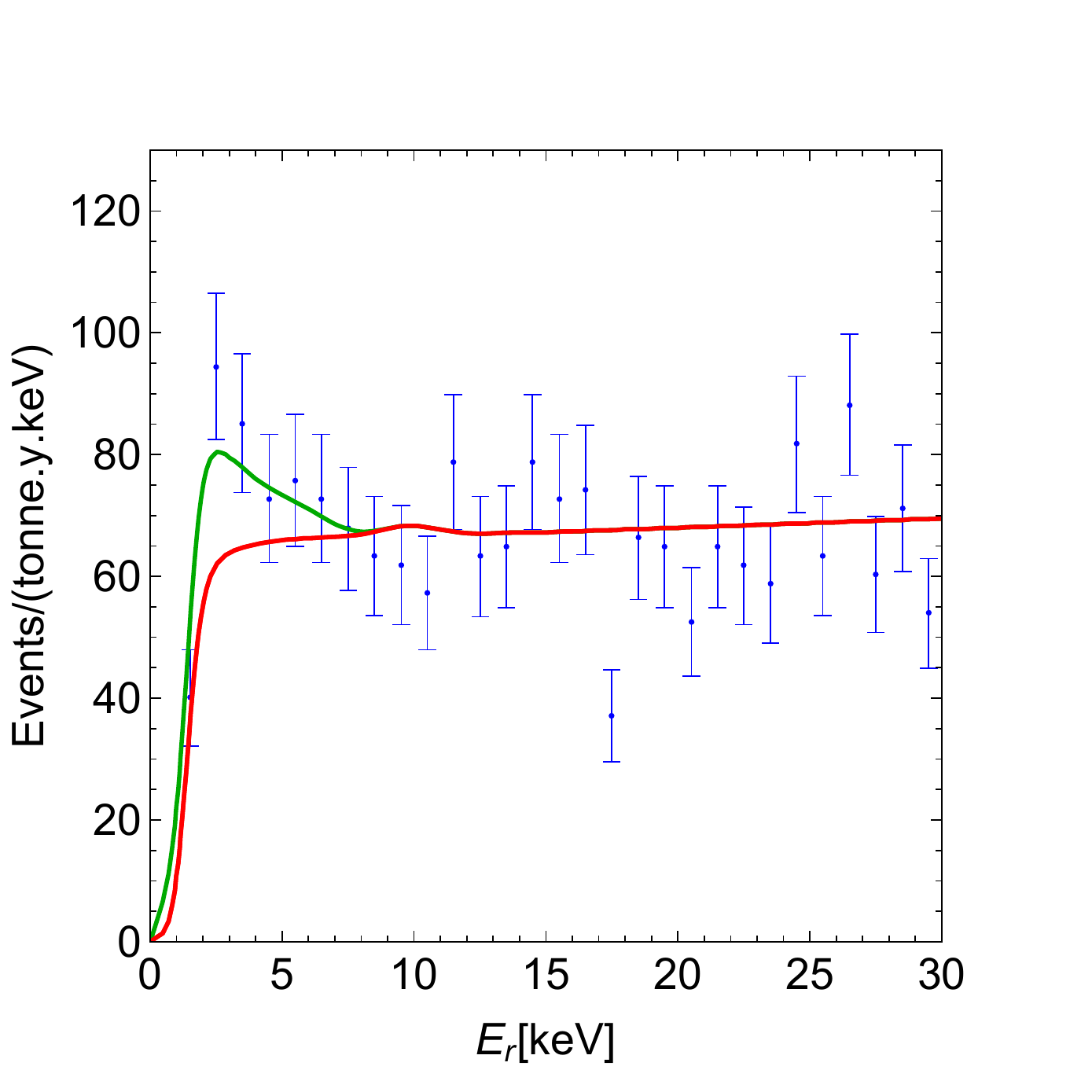}   (c) 
    \includegraphics[width=2.4in]{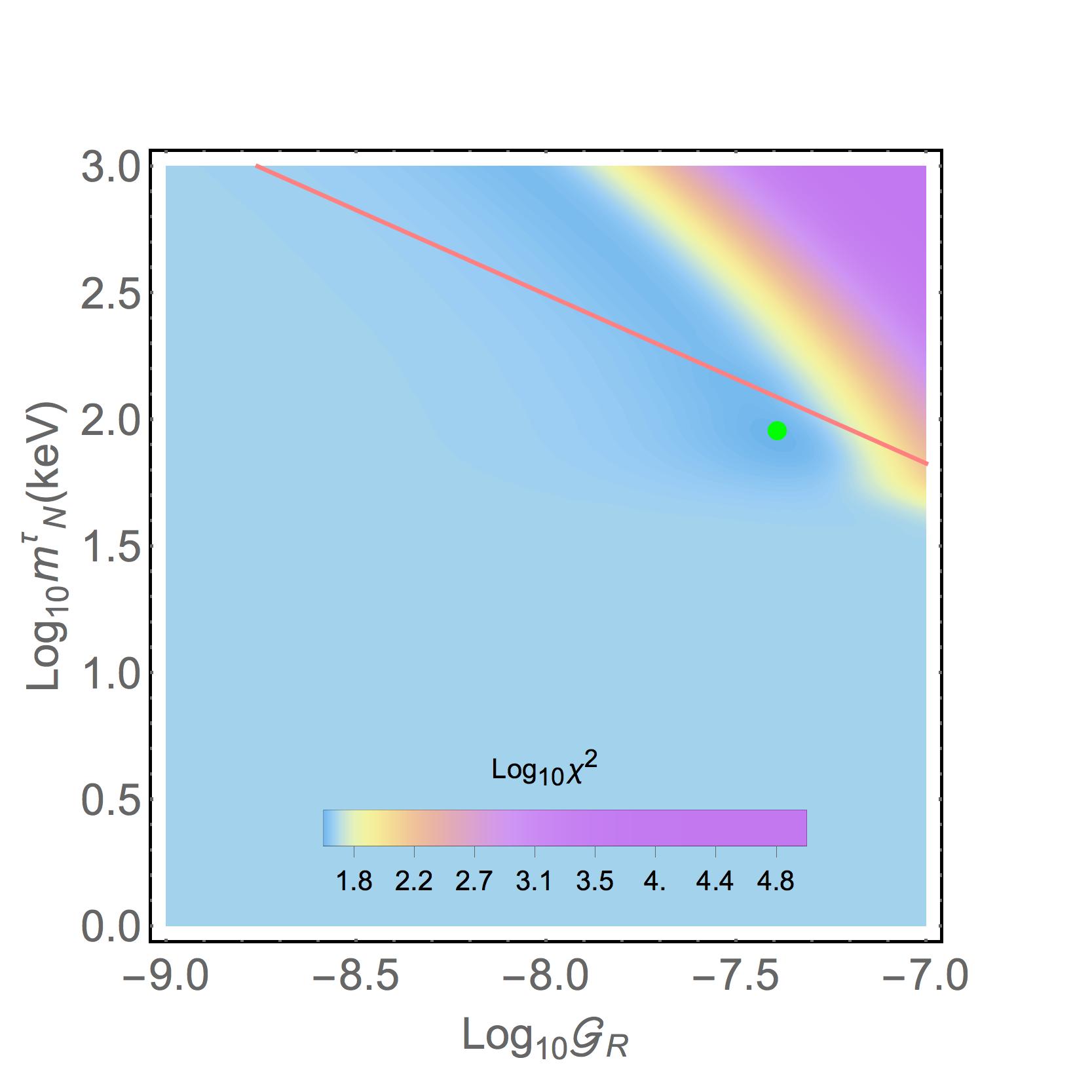}
    \label{fig:event}
    }
    \caption{\small In the left panels events  versus recoil energy including the error bars are shown in blue color. The solid red line is the background model computed by XENON Collaboration \cite{Aprile:2020tmw}. The additional recoil due to the sterile neutrino DM interaction with the Xenon electrons are shown by green solid curve (the best fit points in the right panels are shown by light green points).  To find the minimum  $\chi^2$ we scanned over the coupling  $\mathcal{G}_R$ and the sterile neutrino mass $m^l_N$ considering different charged leptons in the loop (right panels). In case electron (upper panels), muon (middle panels) and tau (lower panels) are in the loop of diagram of Fig.~\ref{enu} the value 
    of coupling and mass of best fit points are 
    $[\mathcal{G}_R,m^l_N]\simeq$ a) $[1.4\times 10^{-4},90.6\text{~keV}]$, b) $[6.4 \times 10^{-7},90.6\text{~keV}]$ and c) $[4.0\times 10^{-8},90.6\text{~keV}]$. The orange lines in right panels are lifetime constraint for sterile neutrinos to be the DM candidate.}
\end{figure}
The value of chi-squared for the fitting parameters used in Figs. \ref{fig:event-electron}, \ref{fig:event-muon} and \ref{fig:event} when we  include all measured data points are $\chi^2|_e\simeq38.3 $, $\chi^2|_\mu \simeq 38.3$ and $\chi^2|_\tau \simeq 38.4$ that implies $\delta \chi|_e\simeq 2.9$, $\delta\chi|_\mu\simeq 2.9$, and $\delta\chi|_\tau\simeq 2.8$, respectively. These results are obtained by considering the first data point around $1$~keV reported by the experiment. If one assumes some deviation from that data point due to the low resolution of detector at low energies, then it is possible to increase the $\mathcal{G}_R$ values to get a better fit by considering the minimum of chi-squared. These values respect the bound from lifetime of sterile neutrinos in Eqs.~(\ref{gRcon}) - (\ref{gRcontau}) as DM candidate. The value of relic abundance does not put a strong bound on the sterile neutrino since to satisfy the relic density constraint one can consider a nonstandard cosmology scenario which was dominated prior to big bang nucleosynthesis \cite{Gelmini:2004ah,Drewes:2013gca}.

The best fit results presented in Figs.~\ref{fig:event-electron} - \ref{fig:event}   weakly depend on the sterile neutrino mass $m^{e,\mu,\tau}_{N}\sim 90$ keV of each family, but $\mathcal{G}_R$ values are rather different and respecting to the DM 
constrains obtained from Eqs. (\ref{gRcon}) - (\ref{gRcontau}). On the basis of these results, we present the following physical interpretations in terms of  $(\mathcal{G}_R,M^{l}_{N})$ values, for massive sterile neutrinos $N^e_R$ ($m^e_{N}$), $N^\mu_R$ ($m^\mu_{N})$ and $N^\tau_R$ ($m^\tau_{N}$):
\begin{enumerate}[(i)]
\item If $\mathcal{G}_R\sim {\mathcal O}(10^{-4})$, only $N^e_R$ is present today as DM component and its mass $m^e_{N}\sim 90$ keV, and $N^\mu_R$ and $N^\tau_R$ have already decayed to  SM particles, e.g., photons and SM neutrinos;

\item  If $\mathcal{G}_R\sim {\mathcal O}(10^{-6})$, sterile neutrinos $N^e_R$ and $N^\mu_R$ are present today as DM particles. $N^\tau_R$ has already decayed to the SM particles. The main contribution to the XENON1T electron recoil 
comes from the $\mu$ channel. This indicates $m^\mu_{N}\sim 90$ keV;

\item  If $\mathcal{G}_R\sim {\mathcal O}(10^{-7})$, all sterile neutrinos $N^e_R, N^\mu_R$, and $N^\tau_R$ are present today as DM particles. The main contribution to the XENON1T electron recoil 
comes from the $\tau$ channel. This indicates $m^\tau_{N}\sim 90$ keV.
\end{enumerate}
These cases are realized by assuming the mass hierarchy $m^e_{N}\lesssim m^\mu_{N}\lesssim m^\tau_{N}$ and also neglecting family mixing phenomenon. 
Regarding the XENON1T electron recoil data, there is a degeneracy between  three scenarios. In order to break the degeneracy and   figure out which situation was selected by the nature, we need other independent experiments/observations to further constrain/determine the values of  $\mathcal{G}_R$ and $m^l_N$.

\subsection{Outgoing Sterile Neutrino}
When the SM neutrino inelastically scattered off the electron producing the sterile neutrinos, the number of events per electron recoil energy $E_{rec}$, per unit detector mass as observed by XENON1T reads %
\begin{eqnarray}
\label{outgevent}
\frac{dR}{dE_{rec}}=N_T \, \int dE_r \int_{E_{\nu}|_\text{min}} dE_{\nu} \frac{d\Phi_{\nu}}{dE_{\nu}} \, \mathcal{P}_{\nu} \, \frac{d\sigma_{tot}}{dE_r} \, \epsilon (E_r) \, G(E_{rec},E_r)\,.
\end{eqnarray}
Here $N_T$ is the number of target electrons in the detector per unit mass, $d\Phi_{\nu}/dE_{\nu}$ is the  low-energy solar neutrino differential flux on Earth  per the incoming neutrino energy $E_{\nu}$ taken from \cite{Bahcall:2004mz,Billard:2013qya}. The minimum energy of neutrinos to produce the recoil energy is denoted by $E_\nu|_{min}$. The parameter $\mathcal{P}_{\nu}$ is the survival probability of one family of neutrinos (e.g.  electron neutrino) arriving at Earth including matter effects and conversion from the two other neutrino families \cite{Lopes:2013nfa,Robertson:2012ib}, $d\sigma_{tot}/dE_{r}$ is the analogous form of Eq.~\eqref{dsigmatotal} for neutrino-electron scattering (see Appendix \ref{appest} for the rough estimation of number of events).  Using Eq.~(\ref{outgevent}) and differential cross sections $\nu e \rightarrow N e$  at tree level in Eq.~(\ref{outgocross}) and its EM channel obtained from the Feynman diagram similar to Fig.~\ref{enu} and Eq.~(\ref{crossloop}) with exchanging sterile and SM neutrinos. Suppose that $\mathcal{G}_{R}$ value preserves SM  precision measurements and astrophysical constraints (which will be discussed in the following section), 
one can not get enough excess to explain XENON1T results.

\section{Bounds from Stellar Cooling in Astrophysics}
It is important to check the consistency of beyond SM couplings (\ref{rhc}), and (\ref{effem0}) with observed astrophysical data and to clarify the available parameter space $({\mathcal G}_R,m_N^l)$ required to fit XENON1T excess data points. The right-handed current couplings (\ref{rhc}), and
their induced neutrino and sterile neutrino EM coupling (\ref{effem0}) have the potential to affect on the stellar energy loss through a plasmon decay channel  $\gamma^{\ast}\rightarrow N_{l}\bar \nu_{l}$.
In order to prevent from anomalous cooling during  stellar evolution 
of the sun, red giants (RGs), horizontal branch stars, supernova, white dwarf,  and etc, \cite{Raffelt:1996wa,Raffelt:1994ry,Arceo-Diaz:2015pva}, we need to put some   constraints on the model parameters.

Sterile neutrino can be easily produced only if sterile neutrino mass $m_{N}$ are smaller than the  plasma frequency $\omega_{p}$, instead, when $m_{N}>\omega_{p}$ the stellar cooling driven by  sterile neutrinos is highly suppressed 
due to kinematical reasons. The most strict astrophysical cooling constraint  can be obtained from the RGs  where the core temperature of a typical RG just before helium ignition was estimated as  18 keV \cite{Raffelt:1996wa,Arceo-Diaz:2015pva}. Obviously, this  temperature  is well below our best fit mass in the range $50$ keV$\lesssim m_N\lesssim 100$ keV, causes less important effects on the cooling process and therefore weaker constraint on $\mathcal{G}_{R}$. However, we can estimate an upper bound on $\mathcal{G}_{R}$ using the similar approach of \cite{Shoemaker:2020kji} by defining an effective transition factor from Eq. (\ref{effem0}) and (\ref{effem1}) [see Fig. \ref{enu2}], effective transition dipole moments in active-sterile neutrino mixing have been previously studied in  \cite{Shoemaker:2018vii,Gninenko:2009ks,Gninenko:2010pr,McKeen:2010rx,Masip:2011qb}. The effective transition magnetic moment operator 
$\mu_{\rm eff} \bar\nu_L\sigma_{\mu\nu}N_R F^{\mu\nu}$ 
has been  considered more recently in \cite{Karmakar:2020rbi,Shoemaker:2020kji,Miranda:2020kwy,Brdar:2020quo} in order to explain XENON1T excess. It was shown that while  the  magnetic moment of active neutrino required to explain the XENON1T excess is $\mu_{\nu}\simeq  2\times 10^{-11}\mu_{B}$,  the active-sterile neutrino transition moment needed for the XENON1T excess is  $\mu_{\text{eff}}\simeq 10^{-10}\mu_{B}$ \cite{Shoemaker:2020kji}.

By qualitatively comparing the neutrino and sterile neutrino EM coupling (\ref{effem0}) with 
the effective transition magnetic moment $\mu_{\text{eff}}$, and the decay rate (\ref{decayr}) with the decay rate $\Gamma_{\mu_{\text{eff}}}=\mu^2_{\text{eff}}m_N^3/(16\pi)$ \cite{Coloma_2017}, 
we arrive at an estimation
\begin{align}
\label{coolbound}
\frac{\mu_{\text{eff}}}{\mu_{B}}\sim \frac{G_{F}m_e}{4\sqrt{2}\pi^{2}}
 \mathcal{G}_{R} m_{\ell}\approx 1.87\times 10^{-12}\left(\frac{\mathcal{G}_{R}}{10^{-2}} \right)\left( \frac{m_{\ell}}{1.77 \text{GeV}} \right)\,,
\end{align}
in order to obtain the upper bounds on $\mathcal{G}_{R}$ from astrophysical data that have imposed the constrain on $\mu_{\text{eff}}$.
By setting the upper bound $\mu_{\text{eff}}$  using more stringent  constraint on neutrino magnetic moment 
$\mu_{\nu}\lesssim 2.2\times10^{-12}\mu_{B}$  inferring from the RGs cooling observations \cite{Shoemaker:2020kji,2019arXiv191010568A}, we can specify allowed parameter region of $\mathcal{G}_{R}$ for each types of the sterile neutrinos in order to be consistent with cooling data can be defined as  $(N^{\tau}_R, \mathcal{G}_{R}\lesssim 1.17\times10^{-2})$,  $(N^{\mu}_R, \mathcal{G}_{R}\lesssim 0.19)$ and $(N^{e}_R, \mathcal{G}_{R}\lesssim 40.83)$ 
from  Eq.~(\ref{coolbound}). These bounds are much weaker than the DM bounds
(\ref{gRcon}), (\ref{gRconmu}) and (\ref{gRcontau}) and also the range of $\mathcal{G}_R$ 
values used to fit XENON1T electron recoil excess in Figs.~\ref{fig:event-electron} - \ref{fig:event}. This shows that our results are consistent with stellar cooling data. There are various types of radiative interactions giving contributions to the stellar cooling process \cite{Magill:2018jla}. While some of the astrophysical observations are subject to  large uncertainties
in both the measurement and the model \cite{DeRocco:2020xdt}, a large region in the parameter space $({\mathcal G}_R,m_N)$ preserving DM constraints  (\ref{gRcon}) - (\ref{gRcontau}) is offered 
to be consistent with both the astrophysical observation and XENON1T new results.

%%%%%%%%%%%%%%%%%%%%%%%%%%%%%%%%%%%%%%%%%%%%%%%%%%%%%%%%%%%%%%%%%%%%%%%%%%%%%%%%%%%%%%%%%%%%%%%%%%%%%%%%%%%%%%%%%%%%%%%%%%%%
%%%%%%%%%%%%%%%%%%%%%%%%%%%%%%%%%%%%%%%%%%%%%%%%%%%%%%%%%%%%%%%%%%%%%%%%%%%%%%%%%%%%%%%%%%%%%%%%%%%%%%%%%%%%%%%%%%%%%%%%%%%%
%%%%%%%%%%%%%%%%%%%%%%%%%%%%%%%%%%%%%%%%%%%%%%%%%%%%%%%%%%%%%%%%%%%%%%%%%%%%%%%%%%%%%%%%%%%%%%%%%%%%%%%%%%%%%%%%%%%%%%%%%%%%

\section{Conclusions and Discussion}
\label{sec:conc}
In the original XENON1T paper~\cite{Aprile:2020tmw}, it is argued that solar axions can provide more satisfactory interpretation while the other explanation such as neutrino magnetic moment or even some possible systematic errors such as the  energy spectrum of tritium decay obtained less statistical significant~\cite{Robinson:2020gfu}. In spite of providing more satisfactory statistical significant by solar  axions, the axion couplings required to fit the data are ruled out by several astrophysical bounds on stellar cooling~\cite{2019arXiv191010568A,Viaux:2013lha,Athron:2020maw,Bertolami:2014wua,Ayala:2014pea,Giannotti:2017hny}.  As an important  example it was recently claimed  that the solar axion explanation of the XENON1T excess 
is in strong tension with the RG cooling bounds \cite{DiLuzio:2020jjp}. 

In this paper, we made an attempt to understand the low-energy excess  of electronic recoils in XENON1T experiment regarding the sterile neutrino DM in the keV mass range which can preserve different bounds.
Our scenario utilizes an effective theory that couples the right-handed sterile neutrino  to the SM particles. We investigated two different cases where  sterile neutrinos can interact with  Xenon electrons either as  incoming or as outgoing particles.
We showed that the excess of events can be explained by inelastic scattering of sterile neutrinos from electrons through EM channel of Eq.~(\ref{crossloop}) in Fig.~\ref{enu}, assuming that incoming flux of sterile neutrinos is provided by the local DM density. During this process the mass of the initial sterile neutrino can be converted into kinetic energy of outgoing SM neutrinos and recoil electrons in  final-state. It is shown  in Figs.~\ref{fig:event-electron} - \ref{fig:event}   that the event rate observed by XENON1T can be reconstructed for the three different families of sterile neutrinos with mass $m^{e,\mu,\tau}_{N}\sim 90$ keV and different values of $\mathcal{G}_{R}$ where best-fit points preserving DM constraints (\ref{gRcon}) - (\ref{gRcontau}). In the case of appearing the sterile neutrino at final state, the solar neutrino flux arrives at detector on the Earth and scatters from Xenon electrons, the process can not produce enough recoil event at low energy. This is due to small values  of $\mathcal{G}_R$ limited by SM precision measurements and astrophysical tests in our model, and also less flux of solar neutrinos compare to local density of sterile neutrino DM.

The rich phenomenology of the scenario presented in this article by introducing three massive sterile neutrinos couple to the SM gauge bosons $W^{\pm}$ and $Z^{0}$ through effective right-handed current interaction (\ref{rhc}), and its new effective EM interaction (\ref{effem0}) of  SM neutrinos $\nu^l_L$ and sterile neutrinos $N^l_R$ might  explain a wide range of anomalous observational evidences \cite{Diaz:2019fwt,Adhikari:2016bei,Abazajian:2012ys} besides satisfying essential astrophysical constraints \cite{Raffelt:1996wa}. The main advantages of our model compare to large growing numbers of beyond SM models on neutrino hidden sector, is that besides having minimal numbers of free parameters, it relies only on the fundamental symmetries and particle content of the SM. It is noteworthy to study this 
model to explain the significant excess of electron-like events  detected in LSND and MiniBooNE experiments \cite{Aguilar:2001ty,Aguilar-Arevalo:2018gpe,Bertuzzo:2018itn,Magill:2018jla}. 
The scattering between solar neutrinos and electrons taking place in Borexino experiment \cite{Agostini:2018uly} may also provide further restrictions on $(\mathcal{G}_{R},m^{l}_{N})$ via SM neutrino-sterile neutrino interaction  
\cite{Coloma:2017ppo,Bringmann:2018cvk,Miranda:2019wdy}.
Moreover, in astrophysics the observed $3.5$ keV line from the center of Galaxy \cite{Bulbul:2014sua,Boyarsky:2014ska}, can be considered in the scenario presented in this article. 
We leave more detailed consideration of different anomalies detected in experiments and observations for further investigations in near future. 

Different scenarios to explain XENON1T excess are required to be confirmed by further
investigations and more precise instruments. In addition to explain XENON1T anomaly, the scenario presented in this article has some distinctive features which can be used to distinguish between our scenario and other beyond SM proposals. 
The new effective EM vertex [Fig. \ref{enu2}] should  give non-vanishing corrections to lepton Anomalous Magnetic Moment (AMM), and  CP violating phases in PMNS and $[(U^\ell_R)^\dagger U^\nu_R]$ mixing matrices provide some new contributions to charged lepton Electric Dipole Moments (EDMs). These new contributions could potentially change EDM and AMM lower bounds to be within future experimental sensitivities and more studies are needed in these directions.

In spite of the sterile neutrino-electron scattering through EM  channel (Fig. \ref{enu}), the scattering via charged lepton coupling (left in Fig. \ref{Fig22}) leads to asymmetry between left- and right-handed recoil electrons. It can be shown that  polarized scattering processes such as $N+e_{L/R}\rightarrow\nu+e_{L/R}$ and $N+e_{R}\rightarrow\nu+e_{L}$ are suppressed with respect to $N+e_{L}\rightarrow\nu+e_{R}$. 
In the limit of all particles being massless due to the conservation of angular momentum,  one can easily verify that  $N$ mainly  scatters from $e_{L}$ giving rise  to recoil $e_{R}$. In order to probe this feature experimentally, we may define the  asymmetry quantity ${\mathcal A}=(\sigma_R-\sigma_L)/(\sigma_R+\sigma_L)$, 
where  $\sigma_R$($\sigma_L$) 
is the cross section of sterile neutrinos with electrons in right-handed  (left-handed) helicity state \cite{Xue:2016dpl,Xue:2013fla,Xue:2001he,Wang:2014bba}. Non-vanishing asymmetry (${\mathcal{A}}\not=0$) induced by the charged current  can be served as a peculiar signature of our model. However, we know that the measurement of the asymmetry  would not be easily achieved in practice, especially for small numbers of events.
In upgrade phase of XENON experiments or other upcoming DM experiments, if
an instrument indirectly detecting the polarization of recoil electrons is implemented then the measurement of the asymmetry  could possibly verify or falsify this feature.

In addition to EM channel of the sterile neutrino-electron scattering  (\ref{s323}), our model predicts other scattering channels (\ref{s321}), (\ref{s322}) and  (\ref{s324}) which may potentially produce observable signals but not in the sensitivity range of current XENON1T experiment. Consequently, the next generation of XENON detectors such
as XENONnT \cite{Aprile:2015uzo}, LZ \cite{Akerib:2015cja}, PandaX-II \cite{Cui:2017nnn} and DARWIN \cite{Aalbers:2016jon}
can  further probe the parameters space  $(\mathcal{G}_R,M^{l}_{N})$ in our model, which will be discussed in our future
papers. Thanks to the multiton-year exposure time of upcoming precise experiments, it may shed light on the dark matter and low-energy neutrino physics beyond SM.

%%%%%%%%%%%%%%%%%%%%%%%%%%%%%%%%%%%%%%%%%%%%%%%%%%%%%%%%%%%%%%%%%%%%%%%%%%%%%%%%%%%%%%%%%%%%%%%%%%%%%%%%%%%%%%%%%%%%%%%%%%%%
%%%%%%%%%%%%%%%%%%%%%%%%%%%%%%%%%%%%%%%%%%%%%%%%%%%%%%%%%%%%%%%%%%%%%%%%%%%%%%%%%%%%%%%%%%%%%%%%%%%%%%%%%%%%%%%%%%%%%%%%%%%%
%%%%%%%%%%%%%%%%%%%%%%%%%%%%%%%%%%%%%%%%%%%%%%%%%%%%%%%%%%%%%%%%%%%%%%%%%%%%%%%%%%%%%%%%%%%%%%%%%%%%%%%%%%%%%%%%%%%%%%%%%%%%

\acknowledgments
The authors are very thankful to Rahul Mehra for insightful discussions on various aspects of direct detection experiments. FH thanks Raghuveer Garani for useful discussions. SS is grateful to Razie Pakravan for her  patience and support during the COVID-19 pandemic issue when this work was in progress. The work of FH is supported by the Deutsche Forschungsgemeinschaft (DFG) through the project number 315477589-TRR 211.

%%%%%%%%%%%%%%%%%%%%%%%%%%%%%%%%%%%%%%%%%%%%%%%%%%%%%%%%%%%%%%%%%%%%%%%%%%%%%%%%%%%%%%%%%%%%%%%%%%%%%%%%%%%%%%%%%%%%%%%%%%%%
%%%%%%%%%%%%%%%%%%%%%%%%%%%%%%%%%%%%%%%%%%%%%%%%%%%%%%%%%%%%%%%%%%%%%%%%%%%%%%%%%%%%%%%%%%%%%%%%%%%%%%%%%%%%%%%%%%%%%%%%%%%%
%%%%%%%%%%%%%%%%%%%%%%%%%%%%%%%%%%%%%%%%%%%%%%%%%%%%%%%%%%%%%%%%%%%%%%%%%%%%%%%%%%%%%%%%%%%%%%%%%%%%%%%%%%%%%%%%%%%%%%%%%%%%
\newpage

\appendix
\section{Kinematics of the Electronic Recoil and Cross Section}
\label{appa}
In this appendix we represent the framework for the computation of  neutrino-electron scattering using our proposed effective interaction taking into account the contribution of sterile neutrinos.  The differential cross section can be evaluated from a general formula given by \cite{Peskin:1995ev,Itzykson:1980rh}
\begin{eqnarray}\label{a1}
d\sigma=\frac{(2\pi)^{4}\delta^{4}(p_{1}+p_{2}-k_{1}-k_{2})}{4\left[(p_{1}\cdot p_{2})^{2}-m_{p1}^{2}m_{p2}^{2}\right]^{1/2}}\frac{d^{3}\vec{k}_{1}}{(2\pi)^{3}2E_{k_{1}}}\frac{d^{3}\vec{k}_{2}}{(2\pi)^{3}2E_{k_{2}}}|M|^{2}\,,
\end{eqnarray}
where $p_{1}$ and $p_{2}$ are four momenta of incoming particle and target electron, respectively, and $k_{1}$ and $k_{2}$ are outgoing scattered particle and the recoil electron, respectively.  The absolute value of Feynman amplitude denoted by $|M|$.

\begin{figure}
    \centering
    \includegraphics[width=1.5in]{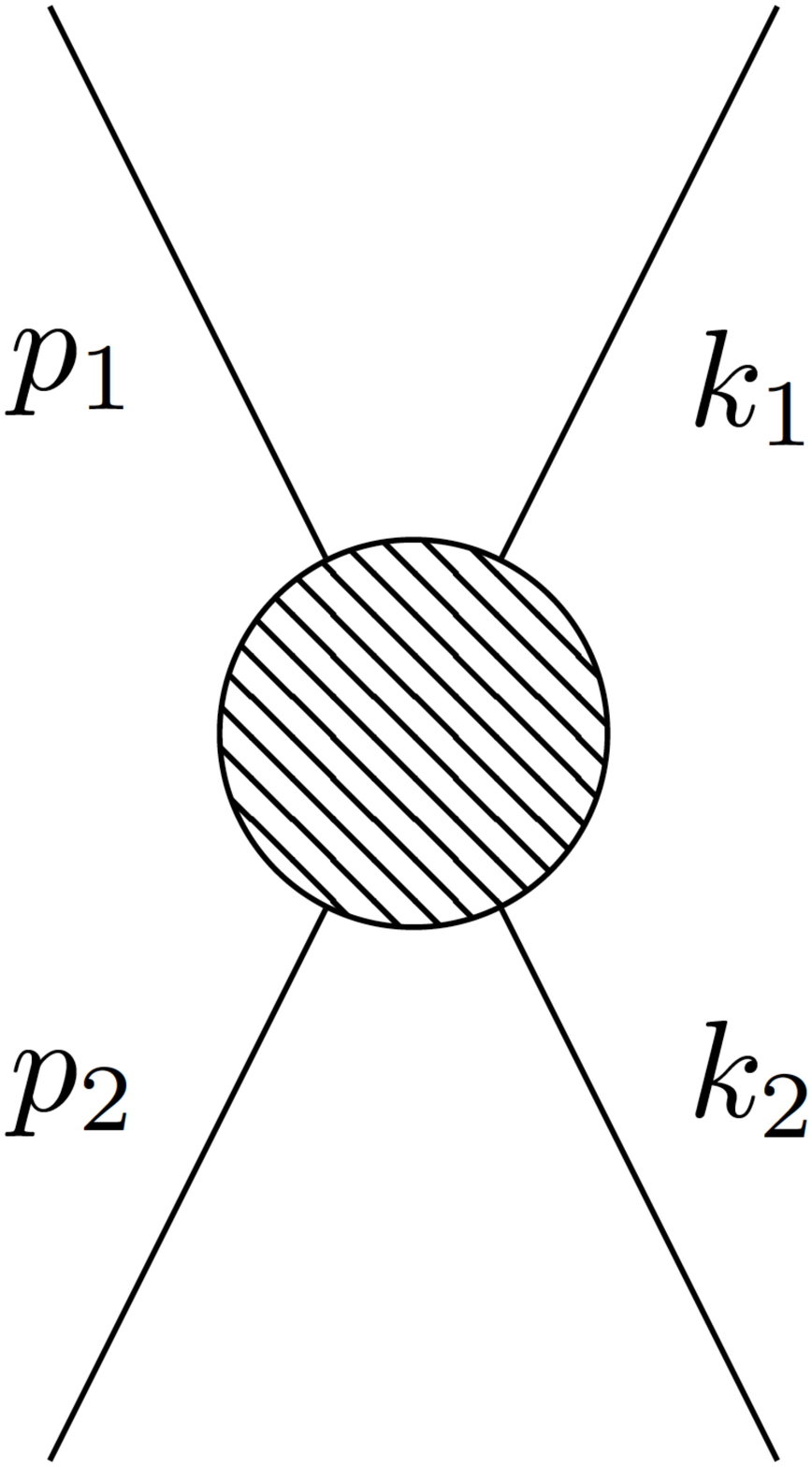}
  \caption{Scattering at a 4-vertex.}
\end{figure}

\subsection*{Incoming Sterile Neutrino}
When modeling direct detection  experiments similar to XENON1T, it is optimal to work
within the Lab reference frame, where ${\vec p}_{2}=0$.
In case $Ne\rightarrow \nu e$ the energy conservation implies $E_{N}+m_{e}=E_{k_{1}}+E_{k_{2}}$, $p_{1}+p_{2}=k_{1}+k_{2}$, $E_{k_{1}}=E_{N}-E_{r}$ and $E_{k_{2}}=m_{e}+E_{r}$. Here 
 $E_r$ is the recoil energy. 
 Then assuming $m_\nu\simeq 0$ we get  
\begin{eqnarray}
\label{incen1}
E_{k_{1}}=\frac{2m_{e}E_{N}+m_{N}^{2}}{2m_{e}+2\left(E_N-\sqrt{E_{N}^{2}-M_{N}^{2}}\cos\Theta\right)}\,,
\end{eqnarray}
for the kinetic energy of the outgoing SM neutrino which implies the minimum energy of nonrelativistic sterile neutrino and the relevant recoil energy to be
\begin{eqnarray}\label{a241}
E_{N}|_{\text{min}} \simeq\frac{E_r}{2} + \frac{1}{2}\sqrt{E_r^2 + 4 m_e E_r+2M_{N}^{2}}\,,
\end{eqnarray}
and
\begin{eqnarray}
E_{r}\simeq\frac{2E_{N}^{2}-M_{N}^{2}}{2(m_{e}+E_{N})}\,.
\end{eqnarray}
The energy of scattered electron is derived as 
\begin{eqnarray}
\label{incen2}
E_{k_{2}}=\sqrt{m_{e}^{2}+E_{k_{1}}^{2}+(E_{N}^{2}-m_{N}^{2})-2E_{k_{1}}\sqrt{E_{N}^{2}-m_{N}^{2}}\cos\Theta} \,.
\end{eqnarray}
The parameter $\Theta$ is the angle between the incoming sterile  neutrino $p_{1}$ and the outgoing neutrino  $k_{1}$, which can be determined by
\begin{eqnarray}
\label{inccos}
\cos\Theta= \frac{1}{\sqrt{E_{N}^{2}-m_{N}^{2}}}\left(
E_{N}-\frac{m_{N}^{2}+2E_{r}m_{e}}{2(E_{N}-E_{r})}
\right)\,.
\end{eqnarray}
Using Eqs. (\ref{a1}), (\ref{incen1}), (\ref{incen2}) and (\ref{inccos}) leads to
\begin{equation}
\label{crossincom}
\frac{d\sigma}{dE_{r}}=
\frac{|M|^{2}}{32\pi m_{e} p_{N}^{2}} \,.
\end{equation}

\subsection*{Outgoing Sterile Neutrinos}
In the case of $\nu_{e} e\rightarrow N e$ we have $E_{\nu}+m_{e}=E_{N}+E_{k_{2}}$ and $\vec p_{\nu}=\vec k_{1}+\vec k_{2}$ from energy momentum conservation. Then by using $E_{k_{2}}=m_{e}+E_{r}$,  $E_{k_{1}}=E_{N}=E_{\nu}-E_{r}$ and $p_{N}=\sqrt{E_{N}^{2}-m_{N}^{2}}$ the following equations for the energy of outgoing electron and the incoming SM neutrino  can be obtained
\begin{eqnarray}
E_{k_{2}}
&=&\sqrt{m_{e}^{2}+E_{\nu}^{2}+(E_{N}^{2}-m_{N}^{2})-2E_{\nu}\sqrt{E_{N}^{2}-m_{N}^{2}}\cos\Theta}\,,
\end{eqnarray}
and 
\begin{eqnarray}\label{28e}
E_{p_{1}}&=&E_{\nu}=\frac{2m_{e}E_{N}-m_{N}^{2}}{2m_{e}-2\left(  E_{N}-\sqrt{E_{N}^{2}-M_{N}^{2}}\cos\Theta\right)}\,.
\end{eqnarray}
The angle between $p_1$ and $k_1$ reads
\begin{eqnarray}\label{29e}
\cos\Theta=1-\left(\frac{m_{N}^{2}+2m_{e}E_{r}}{2E_{\nu}(E_{\nu}-E_{r})}\right)\,.
\end{eqnarray}

The minimum energy of incoming neutrinos that causes the recoil of electron from Eq.~(\ref{28e})
\begin{eqnarray}\label{a24}
E_{\nu}|_{\text{min}} =\frac12 \left(1+\frac{m_N^2}{2m_e E_r}\right)\left(E_r+\sqrt{E_r^2+2m_eE_r}\right)\,,
\end{eqnarray}
and $E_{\nu}|_{\text{max}} = 420 $ keV from the maximum of flux of neutrinos.
Then we can obtain electron recoil energy from (\ref{a24}) in terms of the energy of the incoming SM neutrinos as
\begin{equation}
 \scalebox{0.9}{\text{$E_r= \frac{1}{2m_e(m_e+2E_{\nu})}\left(m_e[2E_{\nu}^2-m_N^2]-m_e m_N^2+E_{\nu}\sqrt{4m_e (E_N^2-m_N^2)-m_N^2(2m_e E_{\nu}-m_N^2)}\right)$}}\,.
\end{equation}
In the limit of $m_{N}\rightarrow 0$ Eqs. (\ref{28e}) and (\ref{29e}) 
correspond to the Compton scattering of photons from electrons in the Lab frame \cite{Peskin:1995ev,Peskin:1995ev}. 
Therefore the differential cross section is 
\begin{equation}
\frac{d\sigma}{dE_{r}}=\frac{1}{\pi m_{e}}\left(\frac{p_{N}}{8 E_{\nu}E_{N}}\right)^{2} \frac{\left[m_{N}^{2}+2m_{e}E_{r} \right]|M|^{2}}{m_{e}(E_{r}+p_{N})+E_{\nu}(p_{N}-E_{N})+m_{N}^{2}}\,.
\end{equation}

\section{Estimation of the Number of Excess Events}
Here we present an order of magnitude estimation of the rate of recoil electrons in the detector for incoming and outgoing sterile neutrinos. The efficiency factor $\epsilon(E_r)$ and the Gaussian smearing function $G(E_r,E_{rec})$ are at the order $\mathcal{O}(1)$ and do not affect the estimation much. 
\subsection*{Incoming Sterile Neutrino}
\label{appest}
Assuming that the sterile neutrino DM inelastically scattered  from the Xenon electrons leading to the SM neutrinos and recoil electrons at the final state, we estimate electronic recoil event rate by using Eq.~(\ref{ingevent}) as
\begin{eqnarray}
\frac{dR_{th}}{dE_{r}}\approx N_T ~ \frac{\rho_{DM}}{m_{N}} \left< f(v) \frac{d\sigma_{N e\rightarrow \nu_{e} e}}{dE_{r}} \right> \approx N_T ~ \frac{\rho_{DM}}{m_{N}}\int dv f(v) v \left(\frac{d\sigma_{Ne}}{dE_{r}}\right)
\,,
\end{eqnarray}
taking the scattering cross section in EM channel  from Eqs.~(\ref{crossloop}) and suppose that  $m_{N}\approx 100$~keV, $E_{N}\simeq m_{N}$,  $v_{N}\approx10^{-3}$ and $p_{N}\simeq m_{N}v_{N}$.
One can find the following approximate value of the differential cross section for recoil energy around a few keV   
\begin{align}
  \frac{d\sigma}{dE_{r}}\approx 5.6\times 10^{-37} \mathcal{G}_{R}^{2}  \left[\frac{1}{\text{cm}^{-2}\text{keV}} \right]\,.
\end{align}
Meanwhile, we assume that the density of sterile neutrinos is obtained from the local density of DM halo as $n_{DM}=(\rho_{DM}/m_{N})\approx 10^{14}\text{cm}^{-2}\text{s}^{-1}$. 
Then our theoretical estimation for the  event rate excess relate to recoil electrons generated via the interaction with sterile neutrinos in a sample of 1 tonne  liquid  XENON  is 
\begin{eqnarray}
\frac{dR_{th}}{dE_{r}}\approx N_T~
\frac{\rho_{DM}}{m_N}~\frac{d\sigma}{dE_{r}} \approx 8.4\times 10^{15} \mathcal{G}_{R}^{2}\left[\frac{1}{\text{tonne. year. keV}}\right]\,,
\end{eqnarray}
regarding the fact that the total number of electrons in XENON target is $N_T\simeq 4\times 10^{27}/$tonne and exposure time of data taking is around $1$ year. 
For the integral over halo velocity distribution we use the result of Refs.~\cite{Drees:2019qzi,Barger:2010gv,DelNobile:2013sia}. 
On the other hand the event  excess that XENON collaboration reported in \cite{Aprile:2020tmw} at low energy is roughly
\begin{align}
\label{exceven}
   \frac{dR_{exp}}{dE_{r}}\approx50  \left[\frac{1}{\text{tonne. year. keV}}\right]\,.
\end{align}
This shows that, for example in the case of incoming tau sterile neutrinos where tau leptons are turning  inside the loop in Fig.~\ref{enu}, by assuming $\mathcal{G}_{R}\approx 7.23 \times 10^{-8}$, one can simply explain this amount of event excess.  Note that this value of 
$\mathcal{G}_{R}$ is consistent with SM precision measurements such as $W^{-}$  decay width and the sterile neutrino lifetime as a DM candidate showing in Eq.~(\ref{decayr}).
\subsection*{Outgoing Sterile Neutrino}
When the sterile neutrino appears in the outgoing state,  using Eq.~(\ref{outgevent}), one can estimate the event rate as 
\begin{eqnarray}
\frac{dR_{th}}{dE_{r}}\approx N_T  \left< \phi_{\nu} \frac{d\sigma_{\nu e \rightarrow N e}}{dE_{r}} \right> \approx N_T \int dE_{\nu} \frac{d\phi_{\nu}}{dE_{\nu}}\left(\frac{d\sigma}{dE_{r}}\right)
\,,
\end{eqnarray}
the incoming neutrino flux is estimated  by the neutrino flux from the proton-proton fusion inside the sun that is around $\phi_{\nu,pp}\approx 6\times10^{10}$~cm$^{-2}\cdot$s$^{-1}$ and the maximum energy of neutrinos is $420$~keV~ \cite{Lopes:2013nfa,Billard:2013qya,Bahcall:2004mz}. 
Then assuming $m_N=100$~keV and using Eqs.~(\ref{outgocross}) and (\ref{dsigmatotal}) for the tree level interaction we can obtain  $d\sigma/dE_{r} \approx 2\times10^{-47}\mathcal{G}_R^2$~[cm$^{2} /$keV]. For the outgoing sterile neutrino case at one loop level (the Feynman diagram similar to Fig.~\ref{enu} and Eq.~(\ref{crossloop}) with exchanging sterile and SM neutrinos) the estimated cross section is 
$d\sigma/dE_{r} \approx 5.6\times10^{-37}\mathcal{G}_R^2$~[cm$^{2} /$keV].
Taking into account the fact that the total number of electrons is $N_T\simeq 4\times 10^{27}/$tonne in addition to the exposure time of the XENON1T experiment \cite{Aprile:2020tmw}. As a consequence, one can provide an explanation for the observed excess ${dR_{exp}}/{dE_{r}}\approx 50~ [\text{tonne. year. keV}]^{-1}$ at a few keV recoil energy, based on the scaled physical parameters which gives 
\begin{eqnarray}
\frac{dR_{th}}{dE_{r}}\approx 8.4\times10^{-1}~\mathcal{G}_R^2 \left[\frac{1}{\text{tonne. year. keV}}\right]\,,
\end{eqnarray}
taking the tree level cross section $\nu e \rightarrow N e$, and 
\begin{eqnarray}
\frac{dR_{th}}{dE_{r}}\approx 4.2\times10^{-4}~\mathcal{G}_R^2 \left[\frac{1}{\text{tonne. year. keV}}\right]\,,
\end{eqnarray}
applying the EM channel scattering cross section $\nu e \rightarrow N e$. These estimations show that to get an excess at the order of  Eq.~(\ref{exceven}), one requires large values for the coupling $\mathcal{G}_R$ because of the small flux of solar neutrinos compare to the DM local density around the earth. These large values of $\mathcal{G}_R$ are in conflict  with SM precision measurements and astrophysical tests. We note that, it is not essential for the outgoing sterile neutrinos to respect  the DM lifetime constraints in Eqs.~(\ref{NRe}) - (\ref{NRtau}). Consequently, in our model the outgoing sterile neutrinos can not explain  the XENON1T excess.

\bibliography{main}

\end{document}